\documentclass{article}
\usepackage[notref,notcite]{}
\usepackage{amsfonts}
\usepackage{amssymb}
\usepackage{amsmath}
\usepackage{graphicx}
\usepackage{slashed}
\usepackage{setspace}
\usepackage{cite}

 \usepackage{color}
 \definecolor{mygray}{gray}{0.45}
 \newcommand{\diagA}[3]{$\!\text{#3\%\,}~{}^{\text{ordinary:\,\,\,{\bf#1}}}_{\text{~~~~super:\,\,\,{\bf#2}}}$\!\!}
 \newcommand{\diagB}[3]{$\!\text{\textcolor{mygray}{#3\%\,}}~{}^{\text{\textcolor{mygray}{ordinary:\,\,\,{\bf#1}}}}_{\text{\textcolor{mygray}{~~~~super:\,\,\,{\bf#2}}}}$\!\!}

\newcommand{\Y}{X}
\newcommand{\coeff}{d}

\def\eqn#1{eq.~(\ref{#1})}

\def\eqns#1#2{eqs.~(\ref{#1}) and~(\ref{#2})}
\def\FMHV{ \cf^{\rm MHV}_n (\eta_{ia}) }
\def\AMHV{ \ca^{\rm MHV}_n }
\def\FNMHV{ \cf^{\rm NMHV}_n (\eta_{ia}) }
\def\tFNMHV{ \tilde\cf^{\rm NMHV}_n (\eta_{ia};\eta_{\Y a})}
\def\tFNMHVf{ \tilde\cf^{\rm NMHV}_5 (\eta_{ia};\eta_{\Y a})}
\def\ANMHV{ \ca^{\rm NMHV}_n}
\def\FNkMHV{ \cf^\text{N$^k$MHV}_n (\eta_{ia})}
\def\tFNkMHV{ \tilde\cf^\text{N$^k$MHV}_n (\eta_{ia};\eta_{\Y a})}
 \def\ANkMHV{ \ca^\text{N$^k$MHV}_n}
\def\bet{ {\bar \eta} }

\newcommand{\reef}[1]{(\ref{#1})}

\newcommand{\cyc}{{\rm cyc}}

\newcommand{\ca}{{\cal A}}

\newcommand{\cn}{{\cal N}}

\newcommand{\cf}{{\cal F}}

\newcommand{\co}{{\cal O}}

\newcommand{\be}{\begin{equation}}
\newcommand{\ee}{\end{equation}}

\def\be{\begin{equation}}
\def\ee{\end{equation}}
\def\bea{\begin{eqnarray}}
\def\eea{\end{eqnarray}}
\def\ba{\begin{array}}
\def\ea{\end{array}}
\def\bd{\begin{displaymath}}
\def\ed{\end{displaymath}}

\def\eg{{\it e.g.~}}
\def\ie{{\it i.e.~}}

\def\a{\alpha}
\def\b{\beta}

\def\d{\delta}

\def\h{\eta}

\def\pa{\partial}

\def\>{\rangle}
\def\<{\langle}
\def\Dsl{D \hskip-.6em \raise1pt\hbox{$ / $ } }
\def\to{\rightarrow}

\def\pa{\partial}

\def\lab{\label}

\newcommand{\eps}{\epsilon}

\def\tQ{\tilde{Q}}

\begin{document}
\bibliographystyle{utphys}

\setstretch{1.05}

\begin{titlepage}

\begin{flushright}
MIT-CTP-4020 \\
BOW-PH-144
\end{flushright}
\vspace{1cm}

\begin{center}
{\Large\bf A super MHV vertex expansion for $\mathbf{\cn =4}$ SYM theory}\\
\vspace{1cm}
{ Michael Kiermaier$^{a}$ ~and~ Stephen G. Naculich${}^{b}$} \\

\vspace{0.7cm}
{{${}^{a}${\it Center for Theoretical Physics}}\\
         {\it Massachusetts Institute of Technology}\\
         {\it 77 Massachusetts Avenue}\\
         {\it Cambridge, MA 02139, USA}}\\[5mm]
{{${}^{b}${\it Department of Physics}}\\
         {\it Bowdoin College }\\
         {\it Brunswick, ME 04011, USA}}\\[5mm]
{\small \tt  mkiermai@mit.edu, naculich@bowdoin.edu}
\end{center}
\vskip .3truecm

\begin{abstract}
\noindent

We present a supersymmetric generalization of the MHV vertex expansion
for all tree amplitudes in $\cn =4$  SYM theory.
In addition to the choice of a reference spinor, this
{\it super MHV vertex expansion}
also depends on four reference Grassmann parameters.
We demonstrate that a significant fraction of diagrams in the expansion
vanishes for a judicious choice of these Grassmann parameters,
which simplifies the computation of amplitudes.
Even pure-gluon amplitudes require fewer diagrams than in
the ordinary MHV vertex expansion.

We show that the super MHV vertex expansion
arises from the recursion relation
associated with a {\it holomorphic all-line supershift}.
This is a supersymmetric generalization of the holomorphic all-line shift
recently introduced in arXiv:0811.3624.
We study the large-$z$ behavior of generating functions under these
all-line supershifts, and find that they generically
provide $1/z^k$ falloff at (Next-to)$^k$MHV level.
In the case of anti-MHV generating functions, we find that a careful
choice of shift parameters guarantees a stronger $1/z^{k+4}$ falloff.
These particular all-line supershifts may therefore play an important role in
extending the super MHV vertex expansion to $\cn=8$ supergravity.

\end{abstract}

\end{titlepage}

\setstretch{0.5}
\tableofcontents
\setstretch{1.05}

\newpage
\setcounter{equation}{0}
\section{Introduction}

On-shell tree-level scattering amplitudes
exhibit a simplicity that is not at all evident
from standard Feynman diagram calculations.
This simplicity can be uncovered using recursion relations for on-shell
amplitudes, of which the MHV vertex expansion \cite{Cachazo:2004kj}
and the BCFW recursion relation \cite{Britto:2004ap,Britto:2005fq}
are prominent examples.
Recursion relations express an on-shell amplitude in terms of simpler
on-shell amplitudes with fewer external legs, thus in principle
allowing a recursive computation of arbitrarily complicated tree amplitudes.

The MHV vertex expansion
expresses an amplitude as a sum over diagrams,
each of which is a product of MHV amplitudes
connected by scalar propagators.
The individual diagrams depend on an arbitrarily-chosen reference spinor
$|X]$,  but the sum of diagrams is independent of $|X]$.
The MHV vertex expansion is a convenient method for
computing amplitudes
due to the simplicity of its basic building blocks, the MHV subamplitudes.
The MHV vertex expansion reproduces the correct on-shell tree amplitudes
both in pure Yang-Mills theory \cite{Risager:2005vk}
and in $\cn=4$ SYM theory~\cite{Elvang:2008na,Kiermaier:2008vz}.

Classes of amplitudes of $\cn=4$ SYM theory that are related by supersymmetric
Ward identities \cite{Grisaru:1976vm,Grisaru:1977px,Bidder:2005in}
can be conveniently packaged into \emph{generating
functions} (also called superamplitudes).
A generating function for $n$-point amplitudes
depends not only on the momenta $p_i$ of the
external particles, but also on $4n$
Grassmann variables $\eta_{ia}$.
(Here, $a=1,2,3, 4$ is an $SU(4)$ index.)
Any particular amplitude can be obtained by acting on the
generating function with a corresponding Grassmann differential
operator~\cite{Bianchi:2008pu}.
The generating function for MHV amplitudes was first given in ref.~\cite{Nair:1988bq}.
Beyond the MHV level, a generating function
for $\cn=4$ SYM amplitudes
can
be represented as a sum over terms involving dual superconformal invariants.
At the NMHV level, this was carried out in
ref.~\cite{Drummond:2008vq}.
The generating function for all tree amplitudes
of $\cn=4$ SYM theory
was then obtained in
ref.~\cite{Drummond:2008cr}
by explicitly solving a supersymmetric generalization of the BCFW recursion
relation \cite{NimaParis,Brandhuber:2008pf,ArkaniHamed:2008gz}.
Alternatively, a generating function beyond the MHV level can be written as a sum over the diagrams of the MHV vertex expansion.
At the NMHV level, this form of the generating function was first presented in ref.~\cite{Georgiou:2004by}.
For general tree amplitudes in $\cn=4$ SYM theory, an explicit representation of the generating function
associated with the MHV vertex expansion was derived in ref.~\cite{Kiermaier:2008vz}.

In this paper we present an alternative
representation for $\cn=4$ SYM generating functions, based on
a new recursion relation, the \emph{super MHV vertex expansion}.
The super MHV vertex expansion
is a natural generalization of the ordinary MHV vertex expansion,
and can be derived from a supersymmetry transformation
acting on the latter.
The diagrams of the super MHV vertex expansion depend not only on a
reference spinor $|X]$,
but also on an arbitrarily-chosen set of four
reference Grassmann parameters $\eta_{\Y a}$.
The sum of diagrams is, of course, independent of these
choices of reference parameters.
With the choice $\eta_{\Y a} = 0$,
the super MHV vertex expansion reduces to the ordinary MHV vertex expansion,
but for well-chosen values of $\eta_{\Y a}$, the super MHV
vertex expansion contains significantly fewer diagrams.
It can thus be used to simplify the computation of both generic
and pure-gluon amplitudes.

The computational simplification occurs because each
of the four $\eta_{\Y a}$ can be chosen to eliminate all diagrams from the
super MHV vertex expansion which contain
an internal line with a particular momentum channel.
As the same momentum channel generically occurs in many diagrams, and as
four distinct channels can be eliminated, many diagrams can be
made to vanish in this way.
Surprisingly, even the computation of pure-gluon tree amplitudes,
which coincide with gluon amplitudes of gauge theory with no supersymmetry,
can be simplified using the super MHV vertex expansion.
We illustrate the simplified computation of amplitudes and the
counting of eliminated diagrams in examples.\footnote{
For a quantitative comparison of the number of diagrams in the
super MHV vertex expansion to the ordinary MHV vertex expansion, the
reader is referred to table~\ref{tabeliminate} in section~\ref{secsimpleNkMHV}.}

Recursion relations can be derived from the analytic behavior
of an on-shell amplitude under a complex shift of its external momenta,
and different shifts generically lead to different
recursion relations\cite{Britto:2005fq,Risager:2005vk}.
Recursion relations were initially developed using
shifts of a subset of the external momenta.
However, the ordinary MHV vertex expansion most naturally follows from a
holomorphic all-line shift,
which deforms all the external momenta \cite{Kiermaier:2008vz}.

We show that the super MHV vertex expansion introduced in this paper follows
naturally from the behavior of generating functions under a {\it holomorphic
all-line supershift}.
This supershift is a generalization of the holomorphic all-line shift
presented in ref.~\cite{Kiermaier:2008vz}, and it shifts not only the
external momenta but also the Grassmann variables $\eta_{ia}$ appearing in
the generating functions.
This is analogous to the generalization of the BCFW two-line shift to the
two-line supershift recently introduced
in ref.~\cite{NimaParis,Brandhuber:2008pf,ArkaniHamed:2008gz}
and applied in ref.~\cite{Drummond:2008cr}.
We prove that $\cn=4$ SYM generating functions beyond the MHV level vanish
when the complex shift parameter $z$ of the all-line supershift becomes large.
This supershift therefore yields a valid recursion relation,
which in turn generates the super MHV vertex expansion.
As in the super BCFW recursion relation, which follows from the two-line
supershift, the diagrams in the super MHV vertex expansion of an amplitude
do not have an immediate interpretation as products of  ordinary tree
amplitudes.  Specific amplitudes, however,
are readily computed directly from the generating function.

We examine the behavior of anti-MHV generating functions
under holomorphic all-line supershifts in detail,
and find that the falloff at large $z$ is faster when the choices
of reference spinor $|X]$ and Grassmann parameters $\eta_{\Y a}$ are
correlated in a certain way.
The analysis of this special case is motivated by a possible future
application of all-line supershift recursion relations
to $\cn=8$ supergravity.
In fact, as we will argue, the extra suppression implies that a
super MHV vertex expansion for $\cn=8$ supergravity
must exist at least for all anti-MHV amplitudes.

\medskip

 This paper is organized as follows. In section~\ref{secreview}, we review MHV amplitudes and the concept of generating functions in $\cn=4$ SYM theory.
 In section~\ref{secsuperMHV}, we present the super MHV vertex expansion as a generalization of the ordinary MHV vertex expansion. We discuss its properties and demonstrate how it can simplify the computation of amplitudes.
 We introduce holomorphic all-line supershifts in section~\ref{secallline}. We study the large $z$ behavior of generating functions under these shifts both for generic and special choices of shift parameters. In section~\ref{secsuperfromall}, we show that the super MHV vertex expansion arises from the recursion relations associated with  holomorphic all-line supershifts. Finally, in section~\ref{secdiscussion}, we discuss the relation of our work to other recent developments. We also comment on the prospects of generalizing the super MHV vertex expansion to $\cn=8$ supergravity.

\setcounter{equation}{0}
\section{Review}\label{secreview}
The simplest on-shell amplitudes in $\cn=4$ SYM
theory
are $n$-point MHV amplitudes.
The MHV sector contains amplitudes with negative helicity gluons on two lines and positive helicity gluons on the remaining lines, together with all amplitudes related to these by supersymmetry. An $n$-point MHV amplitude
takes the simple form\footnote{Throughout this paper, we use the spinor-helicity formalism, with the conventions summarized in
 appendix A of ref.~\cite{Bianchi:2008pu}.}
 \begin{equation}\label{aMHV}
    \AMHV (1,\ldots,n)~=~\frac{\,\<\ldots\>\,\<\ldots\>\,\<\ldots\>\,\<\ldots\>\,}{\cyc(1,\ldots,n)}
    \,, \qquad\text{ with }\qquad \cyc(1,\ldots,n)=\prod_{i=1}^n\<i,i+1\> \,.
 \end{equation}
 The four angle brackets $\<\ldots\>$ in the numerator depend on the choice of states on the external lines $1,\ldots,n$.
For the
pure-gluon MHV
amplitude with negative helicity gluons on lines $i$ and $j$,
for example, the numerator takes the form $\<ij\>^4$, and we obtain the well-known Parke-Taylor formula~\cite{Parke:1986gb}:
 \begin{equation}\label{AMHVij}
    \AMHV (\ldots \,i^-\!\!\!\ldots \,j^-\!\!\!\ldots)~=~\frac{\<ij\>^4}{\cyc(1,\ldots,n)}\,.
 \end{equation}

All $n$-point MHV amplitudes can be conveniently
encapsulated in the $n$-point MHV generating function (or superamplitude)
\cite{Nair:1988bq}
\be
 \label{fMHV}
 \FMHV
 =\frac{\delta^{(8)}\bigl(\sum_{i=1}^n|i\>\eta_{ia}\bigr)}{\cyc(1,\ldots,n)}
 \qquad\text{with}\qquad
\delta^{(8)}\Bigl(\,\sum_{i=1}^n|i\>\eta_{ia}\Bigr)
 =\frac{1}{2^4}\prod_{a=1}^4\,\sum_{i,j=1}^n\<ij\>\eta_{ia}\eta_{ja}\,.
\ee
 Specific amplitudes may then be extracted from $\FMHV$
 by acting with an eighth-order Grassmann differential operator
 $D^{(8)}$ built from operators
 associated with the external states of the amplitude:
 \be
 \AMHV (1,\ldots,n) = D^{(8)} \FMHV\,.
 \ee
 The Grassmann differential operators associated with a particular choice of external particle on line $i$, in order of increasing helicity,
 are given by~\cite{Bianchi:2008pu}:
\begin{equation}\label{Dops}
\begin{split}
D^{-}_i = \frac{\pa^4}{\pa \h_{i1}\pa \h_{i2}\pa\h_{i3}\h_{i4}}\,,\qquad
D_i^{abc} = \frac{\pa^3}{\pa \h_{ia}\pa \h_{ib}\pa\h_{ic}}\,,\qquad
D_i^{ab}= \frac{\pa^2}{\pa \h_{ia} \pa \h_{ib}}\,,\qquad
D^a_i = \frac{\pa}{\pa \h_{ia}} \,,\qquad
D^{+}_i=1\,.
\end{split}
\end{equation}
 The indices $a,b,c=1,\ldots,4$ on the operators $D_i^{abc}$, $D_i^{ab}$, and $D_i^{a}$ associated with scalars and gluinos are $SU(4)$ indices, and distinct choices of  $SU(4)$ indices correspond to distinct choices of scalars and gluinos.
The
pure-gluon
MHV amplitude with negative helicity gluons on lines $i$ and $j$ given
in \eqn{AMHVij}
is now readily obtained from $\FMHV$ as
 \be
 \label{PT}
 D^{(8)} ~=~ D^{-}_iD^{-}_j~~~~\Longrightarrow~~~~
 \AMHV (\ldots \,i^-\!\!\!\ldots \,j^-\!\!\!\ldots)~=~ D^{-}_iD^{-}_j\FMHV~=~\frac{\<ij\>^4} {\cyc(1,\ldots,n)}\,.
 \ee

All non-MHV $n$-point amplitudes may be classified
as (Next-to)$^k$MHV amplitudes for integer $k$ between $1$ and $n-4$.
\footnote{
The maximal value $k=n-4$ corresponds to anti-MHV amplitudes,
which can be expressed as in \eqn{aMHV}, but with angle
brackets replaced by square brackets.}
The N$^k$MHV sector contains amplitudes with $k+2$ negative helicity gluons
and $n-k-2$ positive helicity gluons, together with all amplitudes
related to these by supersymmetry.
All N$^k$MHV amplitudes may also be packaged into generating functions
$ \FNkMHV $, from which a specific amplitude $\ANkMHV$
may be extracted through
 \be
 \label{ADF}
 \ANkMHV (1,\ldots,n) =D^{(8+4k)}\FNkMHV\,.
 \ee
 The order $8+4k$
 Grassmann  differential operator $D^{(8+4k)}$ is built from a product of operators in~(\ref{Dops}) associated with the states on each external line $i$.

\medskip

\setcounter{equation}{0}
\section{The super MHV vertex expansion}
\label{secsuperMHV}

In this section, we explain the super MHV vertex expansion and explore its consequences.
In section~\ref{secNMHV}, we review the ordinary
MHV vertex expansion for NMHV amplitudes and present the super MHV vertex expansion as its generalization.
We also show that the sum rule recently found in ref.~\cite{Kiermaier:2008vz}
is an immediate consequence of this expansion.
We demonstrate in section~\ref{secsimple} that the
reference Grassmann parameters of the super MHV vertex expansion
can be chosen so as to simplify NMHV amplitude computations.
In section~\ref{secNkMHV} we generalize the super MHV vertex expansion to all tree amplitudes in $\cn=4$ SYM  theory, and in section~\ref{secsimpleNkMHV} we analyze the computational simplifications for general tree amplitudes.

\subsection{NMHV amplitudes}\label{secNMHV}
The ordinary MHV vertex expansion \cite{Cachazo:2004kj}
gives a simple prescription to compute arbitrary tree-level amplitudes in $\cn=4$ SYM  theory.
At  the NMHV level, for example, one is instructed to sum over all
possible diagrams in which an $n$-point
NMHV amplitude ${\cal A}^{\rm NMHV}_n$ can be split into two
MHV subamplitudes $I_1$ and $I_2$,
connected by an internal line of momentum $P_\alpha$
(see figure~\ref{fig:N123MHV}a).
Each diagram is characterized by the subset $\alpha$ of external lines
whose momenta flow into the internal line,
\ie
\begin{equation}
    P_\alpha=\sum_{i\in\alpha} p_i\,.
\end{equation}
In fig.~\ref{fig:N123MHV}a, the set $\alpha$
thus consists of all external lines on
subamplitude $I_1$.\footnote{Note that all external momenta $p_i$ are
outgoing in our conventions, which explains the direction of the
arrow on the internal line $P_\alpha$.}
The sum over diagrams gives  the desired NMHV amplitude:
\begin{equation}\label{ANMHV}
 {\cal A}^{\rm NMHV}_n(1,\ldots,n)
=\sum_{\text{diagrams $\alpha$}}{\cal A}^{\rm MHV}(I_1)\,\frac{1}{P^2_\alpha}\,{\cal A}^{\rm MHV}(I_2)\,.
\end{equation}
\begin{figure}
\begin{center}
 \includegraphics[height=5.5cm]{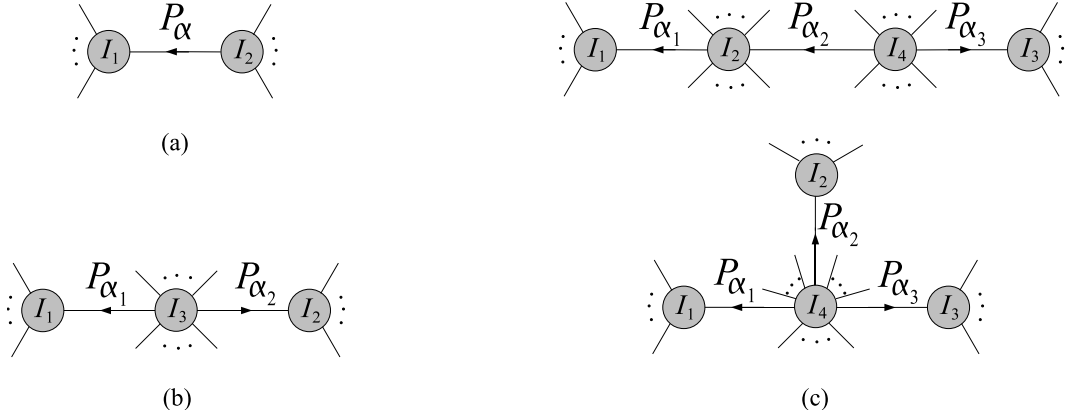}\\[3mm]
\end{center}
\vspace{0mm}
\caption{
The diagrams of the MHV vertex expansion at
the
(a)  NMHV, (b) N$^2$MHV, and (c) N$^3$MHV
level.
}
\lab{fig:N123MHV}
\end{figure}
The momentum $P_\alpha$ of the internal line is not null, and we thus need to explain how to treat the angle brackets $|P_\alpha\>$
which are needed to compute the MHV subamplitudes in \eqn{ANMHV}.
The CSW prescription~\cite{Cachazo:2004kj} instructs us to use
\begin{equation}
    |P_\alpha\>\equiv P_\alpha|X]\,,
\end{equation}
where $|X]$ is an arbitrarily chosen reference spinor. Each diagram in
\eqn{ANMHV} will generically depend on the choice of $|X]$, but their
sum is guaranteed to reproduce the correct amplitude independently of
the chosen reference spinor~\cite{Risager:2005vk,Elvang:2008na}.

All $n$-point NMHV amplitudes
can be packaged into an NMHV generating function,
in terms of which the MHV vertex expansion (\ref{ANMHV}) may
be written as \cite{Georgiou:2004by}
\begin{equation}\label{FnNMHV}
 \cf^{\rm NMHV}_n\bigl(\eta_{ia}\bigr)
 =\sum_{\text{diagrams $\alpha$}}
 \frac{\d^{(8)}\big( \sum_{i=1}^n
 |i\>\h_{ia}\bigr)}{\cyc (I_1)\,P_{\alpha}^2\,\cyc (I_{2})} ~
\prod_{a=1}^4\sum_{i\in \alpha}\<iP_{\alpha}\>\h_{ia}\,.
\end{equation}
Any particular NMHV amplitude can be obtained from this expression
by using \eqn{ADF} with $k=1$.

To define a supersymmetric generalization of \eqn{FnNMHV},
we first recall that supercharges $[Q^a|$ and $|\tQ_a\>$
were defined in~ref.\cite{Bianchi:2008pu} as
\be \lab{schg}
|\tQ_a\> \,\equiv\, \sum_{i=1}^n | i\>\, \h_{ia} \, ,
\hspace{1.5cm} [Q^{a}|  \,\equiv\, \sum_{i=1}^n\, [i|\, \frac{\pa}{\pa \h_{ia}}\,.
\ee
On a function $f(\eta_{ia})$, $[Q^{a}|$ generates the SUSY transformation
\begin{equation}
    f(\eta_{ia})~\longrightarrow~\tilde f(\eta_{ia})~\equiv~\exp\bigl([Q^{a}\,\epsilon_a]\bigr)\,f(\eta_{ia})~=~f\bigl(\eta_{ia}+[\epsilon_a\,i]\bigr)
\end{equation}
and in particular
\begin{equation}\label{shiftFnNMHV}
    \tilde\cf^{\rm NMHV}_n\bigl(\eta_{ia}\bigr)\equiv
\exp\bigl([Q^{a}\,\epsilon_a]\bigr)\,\cf^{\rm NMHV}_n\bigl(\eta_{ia}\bigr)
    =\cf^{\rm NMHV}_n\bigl(\eta_{ia}+[\epsilon_a\,i]\bigr)
\,.
\end{equation}
But since the generating function~(\ref{FnNMHV}) encodes on-shell amplitudes,
it must be SUSY invariant:\footnote{
For the NMHV generating function in the representation~(\ref{FnNMHV}), this was confirmed explicitly in ref.~\cite{Bianchi:2008pu}.}
\begin{equation}\label{SUSYinvariance}
    [Q^{a}\,\epsilon_a]\,\cf^{\rm NMHV}_n(\eta_{ia})=0\,.
\end{equation}
Therefore, although each individual diagram in~(\ref{FnNMHV})
transforms non-trivially under supersymmetry, the function
$\tilde\cf^{\rm NMHV}_n(\eta_{ia}\bigr)$ is
equal to $\cf^{\rm NMHV}_n(\eta_{ia}\bigr) $
after summing over diagrams.
Thus $\tilde\cf^{\rm NMHV}_n(\eta_{ia}\bigr)$
is a valid generating function of
NMHV amplitudes in $\cn=4$ SYM theory.

Let us now choose
\begin{equation}\label{eps}
    [\epsilon_a|=[Y|\eta_{\Y a}\quad\text{ with }\quad [XY]=1\,,
\end{equation}
where we picked the normalization condition for later convenience.
Then
$\tFNMHV$,
defined through \eqns{shiftFnNMHV}{eps},
is the {\it super MHV vertex expansion}
of the NMHV generating function.
Using
\begin{equation}
\begin{split}
    \sum_{i\in \alpha}\<iP_{\alpha}\>\Bigl(\h_{ia}+[Yi]\eta_{\Y a}\Bigr)
    &=\sum_{i\in \alpha}\Bigl(\<iP_{\alpha}\>\h_{ia}+[Yi]\<i|P_{\alpha}|X]\eta_{\Y a}\Bigr)\\
    &=\sum_{i\in \alpha}\<iP_{\alpha}\>\h_{ia}+P_\alpha^2\,\eta_{\Y a}\,,
\end{split}
\end{equation}
the super MHV vertex expansion takes the simple form\footnote{
For $\eta_{\Y a}=0$, the super MHV vertex expansion (\ref{sFnNMHV})
manifestly reduces to the ordinary MHV vertex expansion (\ref{FnNMHV}).}
\begin{equation}\label{sFnNMHV}
\boxed{\phantom{e^{\Bigl(}_{\Bigl(}}\!\!\!
 \tFNMHV
 ~=~\sum_{\text{diagrams $\alpha$}}
 \frac{\d^{(8)}\big(\sum_{i=1}^n|i\>\h_{ia}\bigr)}{\cyc (I_1)\,P_{\alpha}^2\,\cyc (I_{2})} ~
\prod_{a=1}^4\Biggl[\,\sum_{i\in \alpha}\<iP_{\alpha}\>\h_{ia}+P_\alpha^2\,\eta_{\Y a}\Biggr]\,.~
}
\end{equation}
The individual terms of \eqn{sFnNMHV} depend
both on a choice of reference spinor $|X]$
and on a choice of {\it four reference Grassmann parameters} $\eta_{\Y a}$.
For any choice of $|X]$ and $\eta_{\Y a}$, $\tFNMHV$
is the generating function of NMHV amplitudes,
which are obtained,
as in \eqn{ADF},
by applying the twelfth-order Grassmann differential
operator $D^{(12)}$ associated with the
external states of the amplitude:
\be\label{anmhv}
\ANMHV (1,\ldots,n)=D^{(12)}\tFNMHV \,.
\ee
If the Grassmann differential operators in $D^{(12)}$
do not act on $\eta_{\Y a}$, eq.~(\ref{anmhv}) reproduces the
ordinary MHV vertex expansion. However, if we choose
 each $\eta_{\Y a}$ to be a linear combination of the Grassmann
variables $\eta_{1a},\ldots,\eta_{na}$ associated with external states, then
the operators in $D^{(12)}$ do indeed act on $\eta_{\Y a}$.
With such a choice,
$\tFNMHV$ is, diagram by diagram, a sum
of twelfth-order monomials in the Grassmann variables $\eta_{ia}$,
each monomial containing three powers of $\eta_{ia}$
for each fixed $SU(4)$ index $a=1,2,3,4$.
While~\eqn{SUSYinvariance} guarantees the equivalence of
$\tFNMHV$ and $\FNMHV$,
they lead to distinct diagrammatic expansions for amplitudes.
Therefore the super MHV vertex expansion is a non-trivial
generalization of the ordinary MHV vertex expansion when
each $\eta_{\Y a}$ is chosen as some linear combination of the external $\eta_{ia}$.

As $\tFNMHV$  is independent of
$\eta_{\Y a}$
 after summing over diagrams,
its derivative with respect to each $\eta_{\Y a}$
must vanish. We can thus derive sum rules from \eqn{sFnNMHV} by
differentiation. A particularly interesting sum rule can be
obtained by differentiating off the entire
 dependence of $\tFNMHV$ on  $\eta_{\Y a}$,
  yielding
\begin{equation}
 0~=~\biggl[\,\prod_{a=1}^4\frac{\pa}{\pa \eta_{\Y a}} \biggr] \tFNMHV
 ~=~\sum_{\text{diagrams $\alpha$}}
 \frac{\d^{(8)}\big(\sum_{i=1}^n|i\>\h_{ia}\bigr)}{\cyc (I_1)\,P_{\alpha}^2\,\cyc (I_{2})} ~\bigl(P_\alpha^2\bigr)^4\,.
\end{equation}
Pulling out an overall diagram-independent factor of
$\delta^{(8)}(\sum_i|i\>\h_{ia})/\cyc(1,\ldots,n)$, we obtain
\begin{equation}
 0~=\sum_{\text{diagrams $\alpha$}} W_\alpha \bigl(P_\alpha^2\bigr)^4
 ~~~~~\text{with }~~~
 W_\alpha\equiv\frac{\cyc(1,\ldots,n)}{\cyc (I_1)\,P_{\alpha}^2\,\cyc (I_{2})} \,.
\end{equation}
This is the main sum rule derived in section 8.4
of ref.~\cite{Kiermaier:2008vz},
which now finds a natural interpretation as an immediate consequence of the
 $\eta_{\Y a}$  independence of the super MHV vertex expansion.

\subsection{Simplified NMHV amplitude computations}\label{secsimple}
At first sight, the $\eta_{\Y a}$ dependence
of the super MHV vertex expansion~(\ref{sFnNMHV}) may seem
like an unnecessary complication.
This would suggest that the choice
$\eta_{\Y a}=0$, which reduces it to the ordinary MHV vertex expansion,
is most convenient. However, inspection of \eqn{sFnNMHV}
shows that
we can choose the $\eta_{\Y a}$ in such a way that
 certain
diagrams in
the generating function vanish identically.
For example,
pick any
{\it four} diagrams in the MHV vertex expansion and denote them by
$\beta_1,\beta_2,\beta_3$, and $\beta_4$. By choosing
\begin{equation}\label{etaX}
\boxed{\phantom{\Biggl(}
    \eta_{\Y a}~=~-\frac{1}{P_{\beta_a}^2}~\sum_{i\in \beta_a}\<iP_{\beta_a}\>\h_{ia}\,,
    ~ }
\end{equation}
we  guarantee
that the four diagrams $\beta_a$ no longer contribute to the sum over $\alpha$ in the generating function $\tFNMHV$.
Note that this implies that these four diagrams do not contribute to
{\it any} amplitude
$\ANMHV$
computed from \eqn{sFnNMHV}.
We have chosen each $\eta_{\Y a}$ as a different linear combination of the $\eta_{ia}$, to maximize the simplification. We obtain
\begin{equation}\label{sFnNMHV2}
 \tFNMHV
 ~=\!\!\!\!\!\sum_{
\text{diagrams} \atop \alpha\neq \beta_1,\beta_2,\beta_3,\beta_4}
 \frac{\d^{(8)}\big(\sum_{i=1}^n|i\>\h_{ia}\bigr)}{\cyc (I_1)\,P_{\alpha}^2\,\cyc (I_{2})} ~
\prod_{a=1}^4\Biggl[\,\sum_{i\in \alpha}\<iP_{\alpha}\>\h_{ia}-\frac{P_\alpha^2}{P_{\beta_a}^2}~\sum_{i\in \beta_a}\<iP_{\beta_a}\>\h_{ia}\Biggr]\,.
\end{equation}
Here, we have explicitly excluded the diagrams $\beta_1,\ldots,\beta_4$
in the sum, as  diagrams with $\alpha=\beta_a$ involve a
vanishing factor in the product over $a$ and therefore do not contribute.
At the NMHV level, the super MHV vertex expansion thus allows us to eliminate four diagrams.
As we will see below,
many more diagrams can be eliminated at higher N$^k$MHV level. The
simplification thus grows with increasing level in $k$. For now,
let us illustrate the power of the super MHV vertex expansion in the
form~(\ref{sFnNMHV2}) with
an example.

\subsubsection*{Example: NMHV $5$-point amplitudes}
As the simplest example, let us consider  the NMHV amplitude with $n=5$
external lines. (Five-point NMHV amplitudes can be treated as anti-MHV,
which provides a consistency check on our calculation.
Note that
it is our goal to illustrate the advantages of the super MHV vertex
expansion over the ordinary MHV vertex expansion with this example;
we do not expect to achieve simplifications compared to the trivial
anti-MHV computation.) Five diagrams contribute to the ordinary MHV vertex
expansion  for the NMHV 5-point amplitude
(see figure~\ref{fig:NMHV5}a).
\begin{figure}
\begin{center}
 \includegraphics[height=5.5cm]{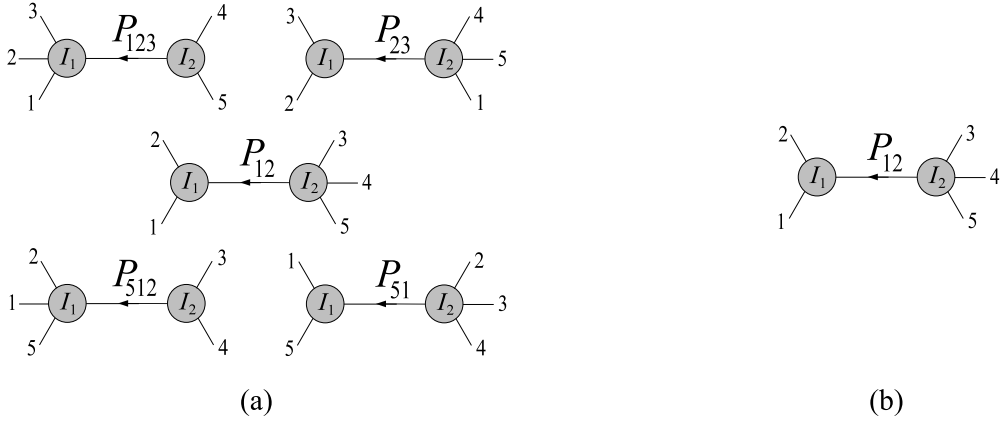}\\[3mm]
\end{center}
\vspace{0mm}
\caption{
(a)~\,The five diagrams contributing to the ordinary MHV vertex expansion of $5$-point NMHV amplitudes.
~~(b)~\,The single remaining diagram that contributes to the super MHV vertex expansion.
}
\lab{fig:NMHV5}
\end{figure}
With the choice (\ref{etaX}) for $\eta_{\Y a}$,
we can eliminate four of
these from the super MHV vertex expansion.
For definiteness, we choose
\begin{equation}\label{4betas}
    \beta_{1}=\{1,2,3\}\,,\qquad \beta_{2}=\{2,3\}\,,\qquad \beta_{3}=\{5,1\}\,,\qquad \beta_{4}=\{5,1,2\}\,.
\end{equation}
The unique diagram which then  contributes
to $\tFNMHVf$
is the diagram $\alpha=\{1,2\}$
(see figure~\ref{fig:N123MHV}b).
Explicitly,
 \eqn{sFnNMHV2} gives
\begin{equation}\label{sFnNMHV5}
 \tilde{\cf}^{\rm NMHV}_5
 ~=~
 \frac{\,\d^{(8)}\big(\sum_{i=1}^n|i\>\h_{ia}\bigr)~
 \prod_{a=1}^4\Bigl[\<1P_{12}\>\h_{1a}+\<2P_{12}\>\h_{2a}-\bigl(P_{12}^2/P_{\beta_a}^2\bigr)\sum_{k\in \beta_a}\<kP_{\beta_a}\>\h_{ka}\Bigr]\,
 }{\<12\>\<2P_{12}\>\<P_{12}1\>~\,P_{12}^2\,~\<34\>\<45\>\<5P_{12}\>\<P_{12}3\>\phantom{\bigl(}} ~
\,.
\end{equation}

To test this generating function, let us compute the gluon amplitude
${\cal A}_5(1^+,2^+,3^-,4^-,5^-)$.
 Since this amplitude is anti-MHV,
the  conjugate of the Parke-Taylor formula immediately gives
\begin{equation}\label{antiMHVresult}
    {\cal A}_5(1^+,2^+,3^-,4^-,5^-)=\frac{[12]^4}{[12][23][34][45][51]}\,.
\end{equation}
Computing this amplitude with the ordinary MHV vertex expansion, however, is messy and the simple result~(\ref{antiMHVresult}) is difficult to obtain analytically. Four diagrams contribute to its expansion;
in fact, precisely the four diagrams $\beta_1,\ldots,\beta_4$ listed
in \eqn{4betas}.
In the super MHV vertex expansion the computation simplifies dramatically. Acting with
$D_3^-D_4^-D_5^-$
on \eqn{sFnNMHV5}, we find
\begin{equation}
\begin{split}
 &D_3^-D_4^-D_5^-\tilde{\cf}^{\rm NMHV}_5\\
 &=
 \frac{~\prod_{a=1}^4
 \frac{\pa}{\pa\eta_{3a}}\frac{\pa}{\pa\eta_{4a}}\frac{\pa}{\pa\eta_{5a}}\Bigl[\frac{1}{2}\sum_{i,j=1}^n\<ij\>\eta_{ia}\eta_{ja}\Bigr]
 \Bigl[\<1P_{12}\>\h_{1a}+\<2P_{12}\>\h_{2a}-\bigl(P_{12}^2/P_{\beta_a}^2\bigr)\sum_{k\in \beta_a}\<kP_{\beta_a}\>\h_{ka}\Bigr]~
 }{\<12\>\<2P_{12}\>\<P_{12}1\>~\,P_{12}^2\,~\<34\>\<45\>\<5P_{12}\>\<P_{12}3\>\phantom{\bigl(}}
\,.
\end{split}
\end{equation}
Consider first the factors $a=1$ and $a=2$. In both cases, the derivatives with respect to $\eta_{4a}$ and $\eta_{5a}$
 must
 act on the first factor, and we obtain
\begin{equation}
\begin{split}
 &\frac{\pa}{\pa\eta_{3a}}\frac{\pa}{\pa\eta_{4a}}\frac{\pa}{\pa\eta_{5a}}\Bigl[\frac{1}{2}\sum_{i,j=1}^n\<ij\>\eta_{ia}\eta_{ja}\Bigr]
 \Bigl[\<1P_{12}\>\h_{1a}+\<2P_{12}\>\h_{2a}-\bigl(P_{12}^2/P_{\beta_a}^2\bigr)\sum_{k\in \beta_a}\<kP_{\beta_a}\>\h_{ka}\Bigr]\\
 &\longrightarrow~\<45\>\frac{P_{12}^2}{P_{123}^2}\<3P_{123}\>=\frac{[12]\<12\>\<3P_{12}\>}{[45]} \hskip4.0cm \text{ for $a=1$}\,,\\[.5ex]
 &\longrightarrow~\<45\>\frac{P_{12}^2}{P_{23}^2}\<3P_{23}\>=-\frac{\<45\>[12]\<12\>[2X]}{[23]}=-\frac{\<45\>[12]\<1P_{12}\>}{[23]} \qquad\text{ for $a=2$}\,.
\end{split}
\end{equation}
The remaining cases $a=3$ and $a=4$ follow by relabeling from the cases $a=2$ and $a=1$, respectively. We obtain
\begin{equation}
\begin{split}
 &D_3^-D_4^-D_5^-\tilde{\cal F}^{\rm NMHV}_5\\
 &=\frac{\bigl([12]\<12\>\<3P_{12}\>\,/\,[45]\bigr)\cdot
 \bigl(-\<45\>[12]\<1P_{12}\>\,/\,[23]\bigr)\cdot
 \bigl(-\<34\>[12]\<2P_{12}\>\,/\,[51]\bigr)\cdot
 \bigl([12]\<12\>\<5P_{12}\>\,/\,[34]\bigr)
 }{\<12\>\<2P_{12}\>\<P_{12}1\>~\,P_{12}^2\,~\<34\>\<45\>\<5P_{12}\>\<P_{12}3\>\phantom{\bigl(}}
\\
 &=\frac{[12]^4}{[12][23][34][45][51]}\,.
\end{split}
\end{equation}
We have thus reproduced the  simple anti-MHV result~(\ref{antiMHVresult}).

The simplicity of this computation was not  just a consequence of our particular choice of external states.
In fact, the generating function $\tFNMHVf$ can  be manipulated to explicitly obtain the anti-MHV generating function.
For example, the $a=2$ factor in \eqn{sFnNMHV5} gives
 \begin{equation}
 \begin{split}
&\<1P_{12}\>\h_{1a}+\<2P_{12}\>\h_{2a}-\frac{P_{12}^2}{P_{23}^2}\Bigl(\<2P_{23}\>\h_{2a}+\<3P_{23}\>\h_{3a}\Bigr)\\
&=\frac{1}{[23]}\Bigl([23]\<1P_{12}\>\eta_{1a}-\<12\>\bigl([23][1X]+[12][3X]\bigr)\eta_{2a}+[12]\<12\>[2X]\eta_{3a}\Bigr)\\
&=\frac{\<1P_{12}\>}{[23]}\Bigl([23]\eta_{1a}+[31]\eta_{2a}+[12]\eta_{3a}\Bigr)\,,
\end{split}
\end{equation}
 where we used the Schouten identity $|1][23]+\text{cyclic}=0\,$.
The other terms can be treated analogously.
The resulting prefactors $\<i P_{12}\>$ cancel the four
$|X]$-dependent
angle brackets in the denominator of \eqn{sFnNMHV5}, and
we obtain
\begin{equation}\label{FantiMHV}
\tilde{\cf}^{\rm NMHV}_5
=\frac{
    \,\d^{(8)}\big(\sum_{i=1}^n|i\>\h_{ia}\bigr)\,}
    {\<45\>^2  \<34\>^2
     \prod_{i=1}^5
    [i,i+1]}\,
\prod_{a=1}^2\Bigl([23]\eta_{1a}+[31]\eta_{2a}+[12]\eta_{3a}\Bigr)
\prod_{a=3}^4\Bigl([25]\eta_{1a}+[51]\eta_{2a}+[12]\eta_{5a}\Bigr)\,,
\end{equation}
which is equivalent to the anti-MHV generating function
for the 5-point amplitude presented in~\cite{Drummond:2008bq,Elvang:2008na}
\begin{equation}\label{FantiMHV2}
    {\cf}^{\overline{\rm MHV}}_5=\frac{
    \,\d^{(8)}\big(\sum_{i=1}^n|i\>\h_{ia}\bigr)\,}
    {\<12\>^4 \prod_{i=1}^5
    [i,i+1]}\,
\prod_{a=1}^4\Bigl([34]\eta_{5a}+[45]\eta_{3a}+[53]\eta_{4a}\Bigr)\,.
\end{equation}
To see this equivalence, we notice that the lines $1$ and $2$ that  are
arbitrarily singled out in~\eqn{FantiMHV2} can in principle be
chosen differently for each value of $a$. In \eqn{FantiMHV},
lines $4$ and $5$ are singled out for $a=1,2\,$, and
lines $3$ and $4$ are singled out for $a=3,4\,$.

\subsubsection*{NMHV amplitudes with $n\geq6$ external states}
For $n\geq6$ external states, more than one diagram contributes to the super MHV vertex expansion of
the NMHV generating function.
For $n=6$ ($n=7$) there are a total of $9$ diagrams ($14$ diagrams) in the ordinary MHV vertex expansion.
As four of these diagrams can be made to vanish by choosing $\eta_{\Y a}$
as in \eqn{etaX},
the super MHV vertex expansion eliminates almost
one-half (one-third)
of the diagrams.
 Many more diagrams can be eliminated for N$^k$MHV amplitudes with $k>1$, and we proceed  to this case now.

\subsection{All tree amplitudes}\label{secNkMHV}
For a general $n$-point (Next-to)$^k$MHV  tree amplitude $\ANkMHV$,
the ordinary MHV vertex expansion instructs us \cite{Cachazo:2004kj}
to sum over all possible
diagrams in which the amplitude  can be split into $k+1$ MHV subamplitudes $I_1$, $I_2$,\ldots, $I_{k+1}$\,, connected by $k$ internal lines of momenta $P_{\alpha_1},\ldots,P_{\alpha_k}$.
(See figures~\ref{fig:N123MHV}b and \ref{fig:N123MHV}c  for the types of MHV vertex diagrams which can occur at N$^2$MHV and N$^3$MHV level, respectively.)
Each diagram is characterized by the  subsets $\alpha_1,\ldots,\alpha_k$ of external lines whose momenta flow into the internal lines,
\ie
\begin{equation}
    P_{\alpha_A}=\sum_{i\in\alpha_A} p_i\,.
\end{equation}
The sum over all possible such MHV vertex diagrams gives the desired N$^k$MHV amplitude:
\begin{equation}\label{ANkMHV}
\ANkMHV (1,\ldots,n)=\!\!\!\!\sum_{\stackrel{\text{MHV diagrams }}{\{\alpha_1,\ldots,\alpha_k\}}}\!
    \frac{{\cal A}^{\rm MHV}(I_1)\cdots {\cal A}^{\rm MHV}(I_{k+1})}{P_{\alpha_1}^2\cdots P_{\alpha_k}^2}\,,
\end{equation}
with the CSW prescription
understood for each occurrence of
the
angle spinors $|P_{\alpha_A}\>$ in
\eqn{ANkMHV}:
\begin{equation}
    |P_{\alpha_A}\>\equiv P_{\alpha_A}|X]\,.
\end{equation}
The generating function associated with the ordinary MHV vertex expansion
is \cite{Kiermaier:2008vz}
\begin{equation}\label{FnNkMHV}
\begin{split}
 &\FNkMHV
 =\sum_{\stackrel{\text{MHV  diagrams }}{\{\alpha_1,\ldots,\alpha_k\}}}
 \frac{ \d^{(8)}\big(\sum_{i=1}^n|i\>\h_{ia}\bigr)}{\cyc (I_1)\cdots\cyc (I_{k+1})} ~
 \prod_{A=1}^k\Biggl[\frac{1}{P_{\alpha_A}^2}\prod_{a=1}^4\sum_{i\in \alpha_A}\<iP_{\alpha_A}\>\h_{ia}\Biggr]\,.
\end{split}
\end{equation}

To obtain the super MHV vertex expansion,
we act with a SUSY transformation
on the generating function~(\ref{FnNkMHV}):
\begin{equation}\label{shiftFnNkMHV}
  \tilde\cf^\text{N$^k$MHV}_n\bigl(\eta_{ia}\bigr)
=\exp\bigl([Q^{a}\,\epsilon_a]\bigr)\,\cf^\text{N$^k$MHV}_n\bigl(\eta_{ia}\bigr)
    =\cf^\text{N$^k$MHV}_n\bigl(\eta_{ia}+[\epsilon_a\,i]\bigr)\,.
\end{equation}
For
the SUSY parameter, we again choose the $[\epsilon_a|$ defined in \eqn{eps}.
The super MHV vertex expansion for the N$^k$MHV generating function then takes the simple form
\begin{equation}\label{sFnNkMHV}
\boxed{\phantom{e^{\biggl(}}\!\!\!\!
 \tFNkMHV
 =\!\!\!\!\sum_{\stackrel{\text{MHV diagrams }}{\{\alpha_1,\ldots,\alpha_k\}}}\!
 \frac{ \d^{(8)}\big(\sum_{i=1}^n|i\>\h_{ia}\bigr)}{\cyc (I_1)\cdots\cyc (I_{k+1})}
 \prod_{A=1}^k\frac{1}{P_{\alpha_A}^2}\prod_{a=1}^4\Biggl[\,\sum_{i\in \alpha_A}\<iP_{\alpha_A}\>\h_{ia}+P^2_{\alpha_A}\eta_{\Y a}\Biggr]\,.~~
}
\end{equation}
Acting on $\tFNkMHV$ with the order $8+4k$  Grassmann differential operator
$D^{(8+4k)}$  associated with an amplitude
$\ANkMHV$, we obtain, diagram by diagram, the super MHV vertex
expansion  for that amplitude.

\subsection{Simplification of general amplitude computations}\label{secsimpleNkMHV}
To simplify general amplitude calculations, we choose the same strategy as
in section~\ref{secsimple} above.
By picking $\eta_{\Y a}$ as in \eqn{etaX} for some choice of channels $\beta_1,\beta_2,\beta_3\,,$ and $\beta_4$, {\it all MHV vertex diagrams for which any internal line $\alpha_A$ coincides with any $\beta_a$ vanish}:
\begin{equation}\label{vanish}
\boxed{\phantom{\Biggl(}
    \alpha_A=\beta_a~~ \text{ for {\it any} }~A=1,\ldots,k\,,~~\text{and {\it any} }~a=1,\ldots, 4
    ~~~~~\Longrightarrow~~~~~\text{diagram vanishes}\,.~}
\end{equation}
For generic amplitude computations\footnote{
If we want to use the super MHV vertex expansion to compute only one specific amplitude, or a specific class of amplitudes (such as pure-gluon amplitudes), a different strategy is advisable. See discussion below.}
the super MHV vertex expansion is most efficient
if we maximize the number of diagrams which vanish by \eqn{vanish}. Different choices of $\beta_a$ can lead to a different number of vanishing diagrams. Two simple guiding principles should be used for the choice of $\beta_a$:
\begin{enumerate}
  \item{\bf The channels $\beta_a$ should appear in as many MHV vertex diagrams as possible.}\\
    Consider $10$-point N$^3$MHV amplitudes. The channel $\alpha=\{1,2,3,4,5\}$ occurs in 123 distinct MHV vertex diagrams,
while the channel $\alpha=\{1,2\}$ occurs in $225$
different diagrams. The choice $\beta_1=\{1,2\}$ is thus more efficient because it eliminates 102
more diagrams than $\beta_1=\{1,2,3,4,5\}$.
  \item {\bf The channels $\beta_a$ should, as far as
  possible, not occur in the same MHV vertex diagrams.}\\
   In the $10$-point N$^3$MHV example, $\beta_1=\{1,2\}$ and $\beta_2=\{2,3\}$ cannot occur together in any MHV vertex diagram.
   The channels
   $\beta_1=\{1,2\}$ and $\beta_2=\{3,4\}$, on the other hand, appear together as internal lines in 20 different MHV vertex diagrams.
In this  case, $\beta_2$ eliminates 20 diagrams that
were already eliminated by $\beta_1$.
The total number of eliminated diagrams is thus reduced
by 20 as compared to the choice $\beta_1=\{1,2\},$ $\beta_2=\{2,3\}$.

   Whether two channels $\beta_1$, $\beta_2$ can occur together in an MHV vertex diagram can be easily tested.
    If, possibly after using the freedom to relabel
    $\beta_1\leftrightarrow \bar\beta_1$ and $\beta_2\leftrightarrow \bar\beta_2$, the sets $\beta_1$ and $\beta_2$ do not share any external lines
    ($\beta_1\cap\beta_2=\emptyset$), then they can appear as internal lines in the same MHV vertex diagram.\footnote{
     Here,
$\bar\beta_a$ denotes
the complement of the set $\beta_a$,
regarded as a subset of all external lines $\{1,\ldots,n\}$. }
\end{enumerate}
\begin{table}[t]
\begin{center}
\renewcommand{\arraystretch}{1.6}
\begin{tabular}{|c|c|c|c|c|c|c|c|c|c|c|}
\hline
&$k=1$ & $k=2$ & $k=3$ & $k=4$ & $k=5$\\
\hline
$n=5$&\diagB{~5}{~1}{80} & & & &\\
\hline
$n=6$&\diagA{~9}{~5}{44} & \diagB{\,~21}{\,~~4}{81}& & &\\
\hline
$n=7$&\diagA{14}{10}{29}& \diagB{\,~56}{\,~23}{59}&\diagB{~~~84}{~~~15}{82}& &\\
\hline
$n=8$&\diagA{20}{16}{20}&\diagA{120}{~\,67}{44}&\diagB{\,~300}{\,~103}{66}&\diagB{~~~330}{~~~~\,57}{83}& \\
\hline
$n=9$&\diagA{27}{23}{15} &\diagA{225}{148}{34} &\diagB{\,~825}{\,~387}{53} &\diagB{\,~1485}{~~~453}{69}& \diagB{~\,1287}{~~~219}{83}\\
\hline
$n=10$&\diagA{35}{31}{11}&\diagA{385}{280}{27}&\diagA{1925}{1085}{44}&\diagB{\,~5005}{\,~2065}{59}&  \diagB{~\,7007}{~\,1967}{72}\\
\hline
$n=11$&\diagA{44}{40}{~\,9} &\diagA{616}{479}{22} & \diagA{4004}{2545}{36} & \diagB{14014}{\,~6989}{50} & \diagB{28028}{10483}{63}\\
\hline
$n=12$&\diagA{54}{50}{~\,7}&\diagA{936}{763}{18}&\diagA{7644}{5285}{31}&\diagA{34398}{19537}{43}&  \diagB{91728}{41447}{55} \\
\hline
$n=13$&\diagA{65}{61}{~\,6}&\diagA{\!\!1365}{\!\!1152}{16}&\diagA{\!\!13650}{\!\!10038}{26}&\diagA{76440}{47712}{38}& \diagB{\!\!259896}{\!\!134316}{48} \\
\hline
$n=14$&\diagA{77}{73}{~\,5}&\diagA{\!\!1925}{\!\!1668}{13}&\diagA{\!\!23100}{\!\!17802}{23}&\diagA{\!\!157080}{\!\!105288}{33}& \diagA{\!\!659736}{\!\!376908}{43} \\
\hline
\end{tabular}
\end{center}
\caption{Comparison of the
\emph{number of diagrams}
in the ordinary MHV vertex expansion and the super MHV vertex expansion for $n$-point N$^k$MHV amplitudes in the range $n=5,\ldots,14$\,,
$k=1,\ldots,5\,$. The \emph{percentage of eliminated diagrams} is also displayed. (Amplitudes which are anti-N$^q$MHV with $q<k$ and thus more efficiently computed using an anti-MHV vertex expansion are displayed in gray.)}
\label{tabeliminate}
\end{table}
In table~\ref{tabeliminate}, we summarize the number of diagrams
eliminated from the generating function for various choices of $n$ and $k$.
For simplicity we always used the choice
 \begin{equation}\label{another4betas}
     \beta_{1}=\{1,2\}\,,\qquad \beta_{2}=\{2,3\}\,,\qquad \beta_{3}=\{3,4\}\,,\qquad \beta_{4}=\{4,5\}
 \end{equation}
for the counting of super MHV vertex diagrams in
table~\ref{tabeliminate}.

The number of diagrams of the ordinary vertex expansion
 at $n$-point N$^k$MHV level 
is given \cite{Roiban:2004yf} by the expression
$M(n,k) = \frac{1}{k+1} {n-3 \choose k} {n+k-1 \choose k}$.
The number of diagrams of the super MHV vertex expansion
for the choice (\ref{another4betas})
is then given by the expression\footnote{We
thank Marcus Spradlin for a question
which prompted us to derive this expression
after v1 of this paper was submitted.}
\begin{equation}
    S(n,k) = M(n,k) - 4 M(n-1,k-1) + 3 M(n-2,k-2)\,.
\end{equation}
Using these analytic expressions for $M(n,k)$ and $S(n,k)$,
the elimination ratio $1-S/M$,
which is also
displayed
in table~\ref{tabeliminate},
can be easily analyzed.
While the percentage of eliminated diagrams decreases with $n$ at fixed $k$,
the elimination ratio increases along the diagonal,
when $k$ and $n$ are increased simultaneously.
Generically, the computationally most challenging amplitudes are
gluon amplitudes with an equal number of negative and positive helicity legs
(and amplitudes related to these by supersymmetry).
For these $2m$-point N$^{(m-2)}$MHV amplitudes, the elimination ratio
remains non-zero
even for $m\to\infty$, approaching
$11/27 \approx 41\%$ in this limit.

\subsubsection*{Pure gluon amplitudes}
If we consider the computation of a particular $\cn=4$ SYM amplitude,
not every diagram contributes to its ordinary MHV vertex expansion.
In fact, there are diagrams for which no assignment
of states to the internal lines can turn all subamplitudes in the diagram
into MHV vertices.
A simple example is the 6-gluon
amplitude ${\cal A}_6^{\rm NMHV}(1^-,2^-,3^-,4^+,5^+,6^+)$, to which
the diagrams $\alpha=\{4,5,6\}$, $\{4,5\}$, and $\{5,6\}$
do not contribute, because the subamplitude containing
the three negative helicity gluon lines $1,2,3$ cannot be MHV.
More generally, each $SU(4)$ index $a$ imposes constraints on the possible diagrams that contribute, and forces certain diagrams to vanish.
For pure-gluon amplitudes, all $SU(4)$ indices appear on the same lines, and they thus all impose the same constraints.
Therefore, many more
MHV vertex diagrams contribute to pure-gluon amplitudes than to generic
$\cn=4$ SYM amplitudes.

In the super MHV vertex expansion, we choose the $\eta_{\Y a}$ to
eliminate diagrams.  Such a choice, however, can cause diagrams to reappear
that would have been absent in the ordinary MHV vertex expansion.
We witnessed this phenomenon in the example of the $5$-point NMHV
amplitude ${\cal A}^{\rm NMHV}_5(1^+,2^+,3^-,4^-,5^-)$
discussed in section~\ref{secsimple}.
There, the only diagram contributing to the super MHV vertex expansion was
$\alpha=\{1,2\}$, precisely the diagram that would not have contributed
to the ordinary MHV vertex expansion.

If we are interested in one particular pure-gluon amplitude, and many diagrams for that amplitude already vanish in its ordinary MHV vertex expansion, it is advisable to pick only three channels $\beta_1,\beta_2,\beta_3$, and to set $\eta_{\Y 4}=0$. The constraints from the $SU(4)$ index $a=4$ then still enforce the vanishing of diagrams absent in its ordinary MHV vertex expansion.

As an example, consider the $8$-point N$^2$MHV amplitude ${\cal A}^{\rm N^2MHV}_8(1^-,2^-,3^-,4^-,5^+,6^+,7^+,8^+)$.
Forty-four
out of $120$ diagrams contribute to its ordinary MHV vertex expansion. We can eliminate
$22$
further diagrams by choosing $\beta_1=\{3,4\}$, $\beta_2=\{3,4,5\}$, and $\beta_3=\{4,5\}$\,. The super MHV vertex expansion of this amplitude thus only
contains
half as many non-vanishing diagrams as the ordinary MHV vertex expansion.
Although we could only use three of the $\beta_a$ to eliminate channels, this elimination ratio of $50\%$ is even better than the generic ratio of $44\%$ given in table~\ref{tabeliminate} for the $8$-point N$^2$MHV level.
Despite its supersymmetric origin, the  super MHV vertex expansion is thus no less powerful when applied to QCD amplitudes.

\setcounter{equation}{0}
\section{All-line supershifts}\label{secallline}

In the next two sections, we will show that the super MHV vertex expansion
presented
above
follows
naturally from the recursion relation
associated with
holomorphic all-line supershifts.
In the current section, we motivate and define holomorphic
all-line supershifts, and study the behavior of generating functions
under these supershifts.

Supershifts were
introduced in ref.~\cite{NimaParis,Brandhuber:2008pf,ArkaniHamed:2008gz}
as a generalization of an ordinary two-line shift in
the BCFW approach \cite{Britto:2005fq}.
An ordinary BCFW shift $[p,q\>$ is defined as
\be
|p] ~\to~ | p] + z |q]\,,
\qquad\qquad
|q\> ~\to~ | q\> - z |p\>\,,
\label{pqshift}
\ee
with all other angle and square spinors remaining unshifted.
Under such a shift of spinors, the scattering amplitude
acquires a dependence on $z$.
If the shift is such that the deformed amplitude $\ca(z)$
vanishes as $z\to\infty$,
then the shift gives rise to a valid
BCFW recursion relation for the amplitude.
In ref.~\cite{Britto:2005fq}, it
was shown that the validity of a shift
depends on the helicities
of the lines $p$ and $q$ participating in the shift.
Consequently, the generating functions, which encode
amplitudes of all helicities, will not generally
vanish at large $z$ under a BCFW shift,
though the coefficients of some of its $\eta$-monomials
will.

A {\it BCFW supershift} $[p,q\>$
is a generalization of \eqn{pqshift}
that acts on Grassmann variables as
well \cite{NimaParis,Brandhuber:2008pf,ArkaniHamed:2008gz,Drummond:2008cr}:
\bea
|p] \to | p] + z |q],
\qquad \qquad
|q\> \to | q\> - z |p\>,
\qquad \qquad
\eta_{pa} ~\to~ \eta_{pa} + z \eta_{qa}\,,
\label{pqsuper}
\eea
with all other angle spinors, square spinors,
and Grassmann variables $\eta_{ia}$ remaining unshifted.
The advantage of the supershift is that
$\sum_{i=1}^n\!|i\>\eta_{ia}$ remains invariant,
which improves the large $z$ falloff of generating functions under this shift.
For example,
the MHV generating function
\be
\FMHV
=\frac{\,\delta^{(8)}\bigl(\sum_{i=1}^n|i\>\eta_{ia}\bigr)\,}
 {\cyc(1,\ldots,n)}
\label{mhvgenfcn}
\ee
vanishes at least as $1/z$ for large $z$, for any choice of lines $p$ and $q$
(it goes as $1/z^2$ if $p$ and $q$ are not adjacent).
This behavior generalizes beyond the MHV level. Indeed,
all generating functions $\FNkMHV$ vanish
at large $z$ for any choice of lines $p$ and $q$
and hence
may be represented by
{\it super BCFW recursion relations}
\cite{NimaParis,Brandhuber:2008pf,ArkaniHamed:2008gz,Drummond:2008cr}.

The MHV  vertex expansion of an amplitude,
on the other hand, may be derived from a holomorphic shift,
\ie a shift
that acts on square spinors only and leaves all angle spinors invariant.
A holomorphic all-line shift
was defined in ref.~\cite{Kiermaier:2008vz} as
\bea
|i] \to | i] + z\,c_i  |X],
\qquad \qquad
|i\> \to | i\>
\qquad\qquad
 \text{for}\quad
i = 1, \ldots, n\,\,,
\label{holo}
\eea
where $|X]$ is an arbitrary reference spinor,
and the complex parameters $c_i$ are constrained by momentum conservation
to obey
\be
\sum_{i=1}^n  c_i |i\>=0\,.
\label{momcons}
\ee
 We also demand that
 the sum of momenta be unchanged under an all-line supershift
 {\it only} when {\it all} external momenta are summed.
 Specifically, we require
 \begin{equation}\label{subsetsumci}
     \sum_{i\in \alpha}c_i|i\>\neq0
 \end{equation}
 for all proper subsets $\alpha$ of consecutive external lines.
The MHV generating function
(\ref{mhvgenfcn})
is manifestly invariant under
this shift.
Furthermore, it was shown in
ref.~\cite{Kiermaier:2008vz} that
N$^k$MHV generating functions
 $\FNkMHV$,
and thus all N$^k$MHV amplitudes, fall
off at least as $1/z^k$ under an all-line shift.
 Therefore all amplitudes in $\cn=4$ SYM theory may be represented by the MHV vertex expansion.

We generalize the shift~(\ref{holo})
to a {\it  holomorphic  all-line supershift}
\bea
 \boxed{ \phantom{\biggl(}
|i] \to | i] + z\,c_i  |X],
\qquad \qquad
|i\> \to | i\>,
\qquad \qquad
\eta_{ia} ~\to~ \eta_{ia} + z\,c_i \eta_{\Y a}
 \qquad\qquad \text{for}\quad i = 1, \ldots, n\,\,,
 ~}
\label{holosuper}
\eea
where, in addition to the reference spinor $|X]$,
we introduce four arbitrary reference Grassmann parameters
$\eta_{\Y a}$.
We still impose the conditions~(\ref{momcons}) and~(\ref{subsetsumci}).
The former in particular implies that
$\delta(\sum_{i=1}^n\!|i\>\eta_{ia})$
is invariant under the supershift~(\ref{holosuper}) since
\begin{equation}\label{invariantdelta}
 \sum_{i=1}^n|i\>\eta_{ia}~~
\to~~\sum_{i=1}^n|i\>\eta_{ia}+z\sum_{i=1}^nc_i|i\>\eta_{\Y a}
=\sum_{i=1}^n|i\>\eta_{ia}\,.
\end{equation}
Consequently, the MHV generating function (\ref{mhvgenfcn})
is invariant under a holomorphic all-line  supershift.
 In section~\ref{secantimhv}, we study the behavior of anti-MHV generating functions under this shift.
In sec.~\ref{seclargez}, we will show that
the N$^k$MHV generating function falls off at least as $1/z^k$
under an all-line supershift,
and in sec.~\ref{secsuperfromall}, we will
use its recursion relation to derive the super MHV vertex expansion.

\subsection{Anti-MHV generating functions}
\label{secantimhv}

We now consider the behavior of anti-MHV generating functions under
supershifts.
Due to their simplicity, anti-MHV generating functions are the ideal testing ground to study the large-$z$ falloff of generating functions under supershifts.
We will examine this falloff
for both
generic holomorphic all-line supershifts and a particularly interesting restricted class of such supershifts.

The $n$-point anti-MHV generating function,
expressed in terms of the conjugate Grassmann variables $\bet_i^a$,
is given by
\be
\overline{\!\cf}^{\,\overline{\rm MHV}}_n
(\bet_i^a)
=\frac{\delta^{(8)}\bigl(\sum_{i=1}^n|i]\bet_i^a\bigr)}
{
\prod_{i=1}^n [i,i+1] }\,.
\label{antigenfcn}
\ee
The numerator of the anti-MHV generating function,
\be
\delta^{(8)} \Bigl(\,\sum_{i=1}^n|i]\bet_i^a\Bigr)
=\frac{1}{2^4} \prod_{a=1}^{4} \sum_{i,j=1}^{n}[i\,j]\,\bet_i^a\,\bet_j^a\,,
\label{delta}
\ee
may be
recast as
a function of $ \eta_{ia}$:
\be
 \delta^{(8)} \Bigl(\sum_{i=1}^n|i]\bet_i^a\Bigr)
 ~~\xrightarrow{\rm GFT}~~
\frac{   \prod_{a=1}^4 \sum_{j_1,\ldots,j_n=1}^n
\eps^{j_1 j_2 \cdots j_n}\, [j_1 j_2]\, \eta_{j_3 a} \cdots \eta_{j_n a} }
{  \left[2\, (n-2)!\right]^4 }\,.
\label{epseta}
\ee
Here, we used the Grassmann Fourier transform
(GFT) \cite{Drummond:2008bq,ArkaniHamed:2008gz}:
\be
{\bar f}( \bet_i^a) ~~\xrightarrow{\rm GFT}~~
{f}(\eta_{ia}) ~\equiv~ \int \prod_{i,a} d\bet_i^a
\,\exp\Bigl(\,\sum_{b,j}\eta_{jb}\bet_j^b\Bigr)\, {\bar f}( \bet_i^a)\,.
\label{fourier}
\ee
Hence the anti-MHV generating function is given by~\cite{Elvang:2008na}
\be
{\cf}^{\overline{\rm MHV}}_n (\eta_{ia})
=
\frac{   \prod_{a=1}^4 \sum_{j_1,\ldots,j_n=1}^n
\eps^{j_1 j_2 \cdots j_n}\, [j_1 j_2]\, \eta_{j_3 a} \cdots \eta_{j_n a} }
{\left[2\, (n-2)!\right]^4\,
\prod_{i=1}^n [i,i+1]} \,.
\label{antimhvgen}
\ee

The sum over
$ \eps^{j_1 j_2 \cdots j_n}\, [j_1 j_2]\, \eta_{j_3 a} \cdots \eta_{j_n a} $
is invariant under a BCFW supershift (\ref{pqsuper}).
Therefore the anti-MHV generating function, just as the MHV-generating
function, falls off as $1/z$ (or $1/z^2$) under a BCFW supershift.

Under a holomorphic all-line supershift (\ref{holosuper}),
the sum over
$ \eps^{j_1 j_2 \cdots j_n}\, [j_1 j_2]\, \eta_{j_3 a} \cdots \eta_{j_n a} $
 generically picks up a piece linear in $z$;
higher powers of $z$ cancel due to the antisymmetry of
$\eps^{j_1 j_2 \cdots j_n}$.
Since each square bracket in the denominator of \eqn{antimhvgen}
generically goes as $z$,
the anti-MHV generating function vanishes as $1/z^{n-4}$
for large $z$ under a holomorphic all-line  supershift (\ref{holosuper}).
Since an $n$-point anti-MHV amplitude is an N$^k$MHV amplitude with $k=n-4$, we conclude that
\begin{equation}\label{MHVfalloff}
{\cf}^{\overline{\rm MHV}}_n (\eta_{ia})~\sim~\frac{1}{z^k}~~
\end{equation}
under a generic holomorphic all-line supershift.

While the falloff~(\ref{MHVfalloff}) is sufficient to derive
all-line supershift recursion relations for anti-MHV generating functions
with $n>4$ external legs,
it is worthwhile to examine whether the $1/z^k$ falloff can be improved
for a careful choice of shift parameters.
This possibility is interesting for two reasons.
First of all, improved large $z$ behavior implies the presence of
additional sum rules.
Secondly, a faster falloff could justify a super MHV vertex expansion
for generating functions in $\cn=8$ supergravity.
The simplicity of the anti-MHV generating functions allows us to
explicitly test the possibility of improved large $z$ behavior.

As we will now see, the anti-MHV generating function indeed falls off
faster than $1/z^{n-4}$
under a restricted class of holomorphic all-line supershifts.
Expand the square spinor $|X]$ and
the Grassmann parameters $\eta_{\Y a}$ in~(\ref{holosuper}) as
\be
|X]  = \sum_{i=1}^n \coeff_i |i],
\qquad\qquad
\eta_{\Y a}  = \sum_{i=1}^n \coeff_i \eta_{ia} \,.
\label{restrict}
\ee
Note that we
use
the same expansion coefficients $\coeff_i$ for both $|X]$ and $\eta_{\Y a}$,
which is a very special non-generic choice.
Under such a supershift, one may show that
\be
\sum_{j_1,\ldots,j_n=1}^n\!\!\!
\eps^{j_1 j_2 \cdots j_n}\, [j_1 j_2]\, \eta_{j_3 a} \cdots \eta_{j_n a}
~~~\to~~~
 \Bigl(1 + z \sum_{i=1}^n c_i \coeff_i\Bigr)
\!\!\sum_{j_1,\ldots,j_n=1}^n\!\!\!
\eps^{j_1 j_2 \cdots j_n}\, [j_1 j_2]\, \eta_{j_3 a} \cdots \eta_{j_n a}  \,.
\ee
We now further restrict the parameters $\coeff_i$ in \eqn{restrict}
to satisfy
\begin{equation}\label{restrict2}
    \sum_{i=1}^n c_i \coeff_i = 0\,,
\end{equation}
making the numerator of \eqn{antimhvgen} \emph{invariant} under the supershift.
Since each square bracket in the denominator of \eqn{antimhvgen} goes as $z$
(for $n>4$),
the anti-MHV generating function vanishes as $1/z^{n}$ at large $z$.
The generic $1/z^k$ behavior
under
a holomorphic all-line supershift
is thus improved, and we have
  \begin{equation}\label{specialfalloff}
    \boxed{\phantom{\Biggl(}
    {\cf}^{\overline{\rm MHV}}_n (\eta_{ia})~\sim~\frac{1}{z^{k+4}}
\text{~~~~~for~~$n>4$,~~~~~when~~}
\sum_{i=1}^n c_i \coeff_i = 0\,. ~~
    }
 \end{equation}
 The significance of this result is that it implies the validity of a super MHV vertex expansion for anti-MHV generating functions in $\cn=8$ supergravity, as we will argue in section~\ref{secdiscussion}.

The improved falloff~(\ref{specialfalloff}) does not
hold
for $n=4$ external legs.
As anti-MHV four-point  functions are also MHV,
they must be invariant under any holomorphic all-line supershift
and cannot
possibly
go as $1/z^4$.
In fact, the kinematics of four-point functions ensures that
the square brackets in the denominator of \eqn{antimhvgen}
are invariant under supershifts that
satisfy the condition~(\ref{restrict2}).
The generating function
for four-point anti-MHV amplitudes
is thus indeed invariant under
such
a supershift.

It is instructive to derive the result~(\ref{specialfalloff}) in a different
way by considering the anti-MHV generating function
in the form (\ref{antigenfcn}).
When condition~(\ref{restrict2}) holds,
the linear shift on $\eta_{i a}$ is equivalent to a linear shift on
$\bar\eta_{i }^a$ in the Fourier-transformed generating function
$\overline{\!\cf}(\bet_i^a)$.
In fact, for any functions ${f}\bigl( \eta_{ia})$
and ${\bar f}\bigl( \bar\eta_i^a)$ which are related by a
Grassmann Fourier transformation,
one has
 \begin{equation}
 \begin{split}
    {f}\bigl( \eta_{ia}+z c_i{\textstyle\sum_j}\,\coeff_j\eta_{ja}\bigr) ~\xrightarrow{\rm GFT}~~&
    \int \prod_{i,a} d\eta_{ia}
    \,\exp\Bigl(\,\sum_{b,j}\bar\eta_j^b\eta_{jb}\Bigr)\, {f}\bigl(\eta_{ia}+z c_i{\textstyle\sum_j}\,\coeff_j\eta_{ja}\bigr)\\
    =&\int \prod_{i,a} d\eta'_{ia}
    \,\exp\Bigl(\,\sum_{b,j}\bigl[\bar\eta_j^b  - z\,\coeff_j \textstyle{\sum_{k}}\, c_k \bar\eta_k^b\bigr]\,\eta'_{jb}\Bigr)\, {f}\bigl(\eta'_{ia}\bigr)
    ~=~\bar f\bigl(\bar\eta_{i}^a  - z\,\coeff_i \textstyle{\sum_j}\, c_j \bar\eta_j^a\bigr)\,.
\end{split}
\end{equation}
Here, the condition~(\ref{restrict2}) was
necessary to show that the change of variables
$ \eta'_{ia} =   \eta_{ia}+z c_i{\textstyle\sum_j}\,\coeff_j\eta_{ja} $
implies
$ \eta_{ia} =   \eta'_{ia}-z c_i{\textstyle\sum_j}\,\coeff_j\eta'_{ja} $
and to guarantee that the measure is invariant under the change of variables.
We conclude that the conjugate Grassmann variables $\bet_i^a$
transform linearly under a supershift:
\be
\boxed{\phantom{\Biggl(}
\eta_{ia} ~\to~  \eta_{ia} + z\, c_i  \sum_{j=1}^n \coeff_j\eta_{ja}
~~~~~\Longleftrightarrow~~~~~~
\bet_i^a  ~\to~  \bet_i^a  - z\,\coeff_i \sum_{j=1}^n c_j \bet_j^a
\qquad\qquad
 \text{when}\quad
\sum_{i=1}^n c_i \coeff_i = 0 \,.
~~}
\ee
Note that the roles of $c_i$ and $\coeff_i$ are reversed
in the shifts of $\eta_{ia}$ and $\bet_i^a$.

Since by \eqn{restrict} we have
\be
|i] \to | i] + z\,c_i  \sum_{j=1}^n \coeff_j |j]\,,
\ee
it follows that
$\sum_{i=1}^n \!|i]\bet_i^a$
and therefore
$\d^{(8)}\big(\sum_{i=1}^{n}\!|i] \bet_i^a\big)$
is
invariant under this restricted class of supershifts.\footnote{
If we impose the even
stronger condition $c_i \coeff_i = 0$ for each $i$,
then the holomorphic all-line supershift is manifestly
a composition of several
$[p,q\>$ supershifts
(\ref{pqsuper})
and the invariance of
$\sum_{i=1}^{n}|i] \bet_i^a$ automatically follows.
Note, though, that such a shift is no longer an all-line shift,
as we need to set at least one $c_i=0$ to satisfy this stronger condition.}
The improved $1/z^n$ falloff is then manifest in the anti-MHV generating
function~(\ref{antigenfcn}).

\subsection{N$^k$MHV generating functions under all-line supershifts}
\label{seclargez}

We now show that generating functions for N$^k$MHV amplitudes fall off at least as $1/z^k$  for large $z$
under all-line supershifts with shift parameters $|X]$ and $\eta_{\Y a}$.
For $k\geq1$ these supershifts thus give
valid recursion relations. As we will show in the following section, the associated recursion relations
imply the super MHV vertex expansion.
Our derivation of the falloff is based on
the ordinary MHV vertex expansion of the generating function,
whose validity was established
in refs.~\cite{Elvang:2008na,Kiermaier:2008vz}.
An alternative derivation,
based on the super BCFW recursion relations of
ref.~\cite{NimaParis,Brandhuber:2008pf,ArkaniHamed:2008gz,Drummond:2008cr}
is outlined in appendix~\ref{appBCFW}.

Consider the behavior of each of the terms in
the ordinary MHV vertex expansion~(\ref{FnNkMHV})
under a generic holomorphic all-line supershift~(\ref{holosuper}).
Crucially, we use a reference spinor $|Z]$ in the ordinary MHV
vertex expansion that does not coincide with the spinor $|X]$ appearing
the supershift (\ref{holosuper}).   We demand $[XZ]\neq0$.
The shift then acts on the CSW spinors $|P_{\alpha_A}\>$ in an MHV vertex diagram as
\begin{equation}
    |P_{\alpha_A}\>=
P_{\alpha_A}|Z]~\to~\hat P_{\alpha_A}|Z]
=P_{\alpha_A}|Z]+z\sum_{j\in \alpha_A}c_j|j\>[XZ]\,.
\end{equation}
Condition~(\ref{subsetsumci}) ensures that the $O(z)$ term
on the right hand side
does not vanish.
For any external line $i$, and any two internal lines $P_{\alpha_A}$ and $P_{\alpha_B}$ of an MHV vertex diagram, we then find the following shift dependence:
\begin{equation}
    \<i\hat P_{\alpha_A}\>\sim z\,, \qquad \<\hat P_{\alpha_A}\hat P_{\alpha_B}\>\sim z^2\,.
\end{equation}
We can thus associate one power of $z$ with each occurrence of $|\hat P_{\alpha_A}\>$.
As the CSW spinor $|\hat P_{\alpha_A}\>$
of each internal line $\alpha_A$
appears four times in the cyclic factors of the denominator
of~(\ref{FnNkMHV}), we have
\begin{equation}
    \frac{1}{\cyc(\hat I_1)\cdots\cyc(\hat I_{k+1})}~\sim~\frac{1}{z^{4k}}\,.
\end{equation}
The spin factors in the numerator of~(\ref{FnNkMHV}) shift as
\begin{equation}
    \sum_{i\in\alpha_A}\<i\hat P_{\alpha_A}\>\hat \eta_{ia}~
    =~z^2\sum_{i,j\in\alpha_A}c_ic_j\<ij\>[XZ]\eta_{\Y a}+O(z)
    ~=~~O(z)\,,
\end{equation}
thus overall the numerator goes at most as $z^{4k}$.
Taking into account the shift dependence of
each of the $k$
propagators,
$\,1/\hat P_{\alpha_A}^2\sim1/z\,$,
we conclude that
the MHV vertex expansion of
the generating function,
diagram by diagram, falls off at least as
\begin{equation}
\boxed{\phantom{\biggl(} \FNkMHV~\sim~\frac{1}{z^k}~~ }
\end{equation}
under holomorphic all-line supershifts.

\setcounter{equation}{0}
\section{The super MHV vertex expansion from all-line supershifts}
\label{secsuperfromall}

In this section we derive
the super MHV vertex expansion from
holomorphic all-line supershifts.
The proof is
a generalization of the derivation of the ordinary MHV vertex expansion
in ref.~\cite{Kiermaier:2008vz}. We will thus emphasize the new aspects of the current proof,
referring the reader to the details in ref.~\cite{Kiermaier:2008vz} for steps that proceed
analogously.

\subsection{All-line supershift recursion relations}

We proved in section~\ref{secallline} that N$^k$MHV generating functions
$\FNkMHV$ with $k\geq1$
vanish as $z \to \infty$ under an all-line supershift.
All-line supershifts can  thus be used to derive a recursion relation for
$\FNkMHV$.
Each diagram in the recursion relation is the product of two
generating functions, connected by a scalar propagator $1/P_\alpha^2$.
We denote the set of external states on one subamplitude by $\alpha$, and their associated Grassmann variables by $\{\eta_{ia}\}_{i\in\alpha}$. Similarly, on the other subamplitude we denote the external states and associated Grassmann variables  by $\bar \alpha$ and $\{\eta_{ia}\}_{i\in\bar\alpha}$, respectively. The
 generating functions of the two subamplitudes
also depend on the Grassmann variable
 $\eta_{P_\alpha a}$
associated with the internal propagator line. In the recursion relation we need to carry out the intermediate state sum by acting with the Grassmann
differential operator
$D^{(4)}_\alpha=\prod_a\pa/\pa\eta_{P_\alpha a}$.
We thus obtain
\begin{equation}\label{alllineRR}
\begin{split}
    \cf^\text{N$^k$MHV}_{n}
    &= \frac{1}{2}\sum_{q=0}^{k-1} \sum_{\alpha}
    D^{(4)}_\alpha
    \frac{
     \,\cf^\text{N$^q$MHV}\bigl(\hat\alpha,-\hat P_\alpha\,;\{\hat\eta_{ia}\}_{i\in \alpha}\,,\eta_{P_\alpha a}\bigr)
     \,\cf^\text{N$^{(k-q-1)}$MHV}\bigl(\hat{\bar\alpha},\hat P_\alpha\,;\{\hat\eta_{ia}\}_{i\in \bar\alpha}\,,\eta_{P_\alpha a}\bigr)
    }{P^2_\alpha}
  \biggr|_{z=z_\alpha}\,.
\end{split}
\end{equation}
The notation  $\hat{\a},~\hat{\bar{\a}},~\hat{P}_\a,\hat\eta_{ia}$ indicates that the momenta and Grassmann variables of the subamplitudes are shifted. They are evaluated at $z=z_\alpha$ satisfying the pole condition $\hat P^2_\alpha(z)=0$, \ie
\begin{equation}\lab{zalph}
    z_\alpha=\frac{P_\alpha^2}{\,\,\sum_{i\in\alpha} c_i\<i|P_\alpha|X]\,}\,.
\end{equation}
Condition \reef{subsetsumci} ensures that $z_\alpha$ is always well-defined, and thus all possible diagrams $\alpha$ contribute to the recursion relation.
Since  $\hat{P}_\a $ is null
when $z=z_\alpha$,
we can write
\be
\hat{P}_\a \,=\, |\hat{P}_\a\>[\hat{P}_\a|  ~~\Longrightarrow~~|\hat P_\alpha\>=\frac{ \hat P_\alpha|X]}{[\hat P_\alpha X]}=\frac{ P_\alpha|X]}{[\hat P_\alpha X]}\,.
\label{nullP}
\ee
Finally,
the symmetry factor $\frac{1}{2}$ in \eqn{alllineRR} is necessary because,
for each channel $\alpha$, we now also include the equivalent term
with $\alpha\leftrightarrow\bar \alpha$ in the sum.

\subsection{NMHV generating function}
At the NMHV level, all subamplitudes in the recursion relation~(\ref{alllineRR}) are MHV. Using the MHV generating function~(\ref{fMHV})  we obtain
\begin{equation}
    \cf^\text{NMHV}_{n}    = \sum_{\text{diagrams }\alpha}
    D^{(4)}_\alpha
    \frac{
    \delta^{(8)}\bigl(\sum_{i\in \alpha}|i\>\hat\eta_{ia}-|\hat P_\alpha\>\eta_{P_\alpha a}\bigr)~
    \delta^{(8)}\bigl(\sum_{i\in \bar\alpha}|i\>\hat\eta_{ia}+|\hat P_\alpha\>\eta_{P_\alpha a}\bigr)
}{\cyc(\hat I_1)~P^2_\alpha~\cyc(\hat I_2)}
  \biggr|_{z=z_\alpha}\,.
\end{equation}
where $|\hat P_\alpha\> $ in the numerator,
and in the cyclic factors in the denominator,
is given by \eqn{nullP}.
Using $\delta^{(8)}(A)\delta^{(8)}(B)=\delta^{(8)}(A)\delta^{(8)}(A+B)$
and the invariance of $\sum_{i=1}^n\!|i\>\eta_{ia}$
from
\eqn{invariantdelta},
we obtain
\begin{equation}\label{cfNMHV1}
    \cf^\text{NMHV}_{n}
    = \sum_{\text{diagrams }\alpha}
    \frac{\delta^{(8)}\bigl(\sum_{i=1}^n|i\>\eta_{ia}\bigr)}{\cyc(I_1)~P^2_\alpha~\cyc( I_2)}
    \prod_{a=1}^4\sum_{i\in\alpha}\<i P_{\alpha}\>\hat\eta_{ia}\biggr|_{z=z_\alpha}\,.
\end{equation}
Since all factors of $[\hat{P}_\a\, X]$ cancel, we
used the CSW prescription
\be
|\hat P_\alpha\>~\to~ P_\alpha|X] = | P_\alpha\>
\ee
for all occurrences of $|\hat P_\alpha\>$ in the numerator
and the cyclic factors in the denominator
 of \eqn{cfNMHV1}.

Let us now examine the effect of the shifted $\hat \eta_{ia}$ in \eqn{cfNMHV1}. We find
\begin{equation}
\sum_{i\in\alpha}\<i P_{\alpha}\>\hat\eta_{ia}\biggr|_{z=z_\alpha}
=~\sum_{i\in\alpha}\<i P_{\alpha}\>\eta_{ia}+z_\alpha\sum_{i\in\alpha}c_i\<i P_{\alpha}\>\eta_{\Y a}\
~=~\sum_{i\in\alpha}\<i P_{\alpha}\>\eta_{ia}+P_\alpha^2\,\eta_{\Y a}\,,
\end{equation}
where we
inserted
 $z_\alpha$ from \eqn{zalph} in the last step. We obtain
\begin{equation}
    \cf^\text{NMHV}_{n}
    = \sum_{\text{diagrams }\alpha}
    \frac{\delta^{(8)}\bigl(\sum_{i=1}^n|i\>\eta_{ia}\bigr)}{\cyc(I_1)~P^2_\alpha~\cyc( I_2)}~
    \prod_{a=1}^4\Biggl[\,\sum_{i\in\alpha}\<i P_{\alpha}\>\eta_{ia}+P_\alpha^2\,\eta_{\Y a}\Biggr]\,.
\label{fNMHV2}
\end{equation}
We have thus reproduced
$\tFNMHV$ given in \eqn{sFnNMHV}, \ie the form of the NMHV generating function
associated with the super MHV vertex expansion.

\subsection{N$^2$MHV generating function}
Let us now use \eqn{alllineRR} to determine the generating function at the N$^2$MHV level. One subamplitude is again MHV,
for which we
use the MHV generating function in the form~(\ref{fMHV}). For the NMHV subamplitude,
it is crucial that we use the generating function (\ref{fNMHV2})
associated with the super MHV vertex expansion,
and that we use  the same reference parameters
$|X]$, $\eta_{\Y a}$ in the generating function as in the all-line supershift. The NMHV generating function contains a sum over channels which we denote by $\beta$. The propagator line $P_\alpha$ is an ``external" line of the NMHV subamplitude, but we choose $\b$ to {\it not} include this line ($P_\alpha\notin \beta$), so that both $\alpha$ and $\beta$ only contain external lines of the full amplitude
(see figure~\ref{fig:N2MHVallline}).
\begin{figure}
\begin{center}
 \includegraphics[height=3.5cm]{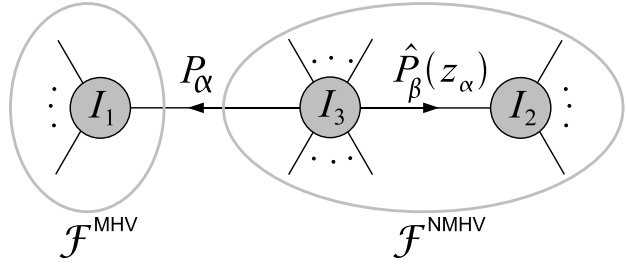}\\[3mm]
\end{center}
\vspace{0mm}
\caption{
A diagram $\alpha$ of the all-line supershift recursion relation of an N$^2$MHV generating function. The super MHV vertex expansion is substituted for the NMHV subamplitude, and the internal line $P_\beta$ of the super MHV vertex diagram is evaluated at shifted momenta.}
\lab{fig:N2MHVallline}
\end{figure}

A short calculation analogous to
that in
ref.~\cite{Kiermaier:2008vz} gives
\begin{equation}\label{fN2MHV1}
    \cf^{\rm N^2MHV}_{n}
    =\sum_{\alpha,\beta}
    \frac{\delta^{(8)}\bigl(\sum_{i=1}^n|i\>\eta_{ia}\bigr)}
    {\cyc ( I_1)\cyc  ( I_2)\cyc ( I_3)}
    \,\frac{1}{P^2_\alpha  \,  \hat P^2_\beta(z_\alpha)}
    \,\prod_{a=1}^4\Biggl[\,\sum_{i\in\alpha}\<i\,P_\alpha\>\hat\eta_{ia}\Biggr]
    \Biggl[\,\sum_{i\in\beta}\<i\,P_\beta\>\hat\eta_{ia}+\hat P_{\beta}^2(z_\alpha)\,\eta_{\Y a}\Biggr]\,.
\end{equation}
Note that the angle brackets $\<i\,P_\beta\>$
(including those implicit in $\cyc(I_2)$ and $\cyc(I_3)$)
are unaffected by the shift because
\be
\label{unaffected}
|\hat P_\beta (z_\alpha) \> =
\hat P_\beta (z_\alpha) |X]  =
 P_\beta |X]  + z_\alpha \sum_{i\in\beta} c_i | i\> [X X]   =
 P_\beta |X]   = | P_\beta \> \,.
\ee
To simplify the last factor in \eqn{fN2MHV1}, we employ
the crucial identity
\begin{equation}\label{crucial}
\boxed{\phantom{\Biggl(}
\sum_{i\in\beta}\<i\,P_\beta\>\hat\eta_{ia}+\hat P_{\beta}^2(z_\alpha)\eta_{\Y a}
=\sum_{i\in\beta}\<i\,P_\beta\>\eta_{ia}+ P_{\beta}^2\,\eta_{\Y a}\,,~}
\end{equation}
where we used
\begin{equation}
z_\alpha\sum_{i\in\beta}\<i\,P_\beta\>+\hat P_{\beta}^2(z_\alpha)=P_\beta^2\,.
\end{equation}
Note that the identity~(\ref{crucial}) relies on the fact that the all-line supershift recursion relation and the NMHV generating function for the subamplitude are based on the same reference parameters $|X]$ and $\eta_{\Y a}$.
We can then rewrite \eqn{fN2MHV1} as
\begin{equation}
\cf^{\rm N^2MHV}_{n}=\sum_{\alpha,\beta}
    \frac{\delta^{(8)}\bigl(\sum_{i=1}^n|i\>\eta_{ia}\bigr)}
    {\cyc ( I_1)\cyc  ( I_2)\cyc ( I_3)}
    \,\frac{1}{P^2_\alpha  \,  \hat P^2_\beta(z_\alpha)}
    \,\prod_{a=1}^4\Biggl[\,\sum_{i\in\alpha}\<i\,P_\alpha\>\eta_{ia}+P_{\alpha}^2\,\eta_{\Y a}\Biggr]
    \Biggl[\,\sum_{i\in\beta}\<i\,P_\beta\>\eta_{ia}+P_{\beta}^2\,\eta_{\Y a}\Biggr]\,.
\end{equation}
Symmetrizing the sum in $\alpha\leftrightarrow\beta$ and using the
identity \cite{BjerrumBohr:2005jr}
\begin{equation}
    \frac{1}{P_\alpha^2\hat P_\beta^2(z_\alpha)}+\frac{1}{\hat P_\alpha^2(z_\beta)\,P_\beta^2}
    =\frac{1}{P_\alpha^2P_\beta^2}\,,
\end{equation}
we find
\begin{equation}\label{fN2MHV2}
\cf^{\rm N^2MHV}_{n}=\frac{1}{2}\sum_{\alpha,\beta}
    \frac{\delta^{(8)}\bigl(\sum_{i=1}^n|i\>\eta_{ia}\bigr)}
    {\cyc ( I_1)\cyc  ( I_2)\cyc ( I_3)}
    \,\frac{1}{P^2_\alpha  \, P_\beta^2}
    \,\prod_{a=1}^4\Biggl[\,\sum_{i\in\alpha}\<i\,P_\alpha\>\eta_{ia}+P_{\alpha}^2\,\eta_{\Y a}\Biggr]
    \Biggl[\,\sum_{i\in\beta}\<i\,P_\beta\>\eta_{ia}+P_{\beta}^2\,\eta_{\Y a}\Biggr]\,.
\end{equation}
As there was no restriction on our original sum over $\alpha$, we are counting each distinct MHV vertex diagram twice in \eqn{fN2MHV2}. The factor of $\frac{1}{2}$ compensates this overcounting. We then have (after relabeling $\alpha\to\alpha_1$ and $\beta\to \alpha_2$)
\begin{equation}
\cf^{\rm N^2MHV}_{n}=\!\!\sum_{\stackrel{\text{MHV diagrams }}{\{\alpha_1,\alpha_2\}}}
    \frac{\delta^{(8)}\bigl(\sum_{i=1}^n|i\>\eta_{ia}\bigr)}
    {\cyc ( I_1)\cyc  ( I_2)\cyc ( I_3)}
    \,\prod_{A=1}^2\,\frac{1}{\,P^2_{\alpha_A}} \,\prod_{a=1}^4\Biggl[\,\sum_{i\in\alpha_A}\<i\,P_{\alpha_A}\>\eta_{ia}+P_{\alpha_A}^2\eta_{\Y a}\Biggr]\,.
\end{equation}
We have thus derived the super MHV vertex expansion~(\ref{sFnNkMHV}) at the N$^2$MHV level.

\subsection{All tree amplitudes}
The generalization to the N$^k$MHV  generating function
with $k\geq3$ is straightforward. We continue inductively in $k$, plugging in the super MHV vertex expansion for the N$^q$MHV subamplitudes ($q<k$) in the recursion relation~(\ref{alllineRR}). We use the same reference parameters
$|X]$, $\eta_{\Y a}$ in the subamplitudes as in the all-line supershift recursion relation. The proof proceeds precisely as
in ref.~\cite{Kiermaier:2008vz}, except that we need to use the identity~(\ref{crucial}) for all channels $\beta_1,\ldots,\beta_q$ appearing in an N$^q$MHV subamplitude. The only remaining shift dependence then
resides in the propagators,
and using the identity~[\citen{BjerrumBohr:2005jr},\citen{Kiermaier:2008vz}]
\begin{equation}
    \sum_{A=1}^k
    \frac{1}{\hat P^2_{\alpha_1}(z_{\alpha_A})\cdots \hat P^2_{\alpha_{\!A-1}}\!(z_{\alpha_A}) \, P^2_{\alpha_{\!A}}\,\hat P^2_{\alpha_{\!A+1}}\!
    (z_{\alpha_A})\cdots \hat P^2_{\alpha_k}\!(z_{\alpha_A})}
    =\frac{1}{P_{\alpha_1}^2\cdots P_{\alpha_k}^2}\,,
\end{equation}
 we obtain the generating function
\begin{equation}
\FNMHV
 =\!\!\!\!\sum_{\stackrel{\text{MHV diagrams }}{\{\alpha_1,\ldots,\alpha_k\}}}\!
 \frac{ \d^{(8)}\big(\sum_{i=1}^n|i\>\h_{ia}\bigr)}{\cyc (I_1)\cdots\cyc (I_{k+1})}
 \prod_{A=1}^k\frac{1}{P_{\alpha_A}^2}\prod_{a=1}^4\Biggl[\,\sum_{i\in \alpha_A}\<iP_{\alpha_A}\>\h_{ia}+P^2_{\alpha_A}\eta_{\Y a}\,\Biggr]\,.
\label{complete}
\end{equation}
This generating function coincides with the super MHV vertex expansion~(\ref{sFnNkMHV}), and completes our derivation.

As a consistency check, one can show that the super MHV vertex expansion
immediately implies the $1/z^k$ falloff of N$^k$MHV
generating functions under all-line supershifts.
This check is carried out in appendix~\ref{appconsistency}.

\setcounter{equation}{0}
\section{Discussion}\label{secdiscussion}

In this paper we  have presented a new family of representations
for the generating functions of $\cn=4$ SYM theory,
the super MHV vertex expansion. The diagrams of this family of
representations depend  on a reference spinor $|X]$
and on four reference Grassmann parameters $\eta_{\Y a}$, which may be
chosen arbitrarily.
 We have shown that the super MHV vertex expansion
 arises both from a particular supersymmetry transformation on the ordinary MHV vertex expansion, and from
 the recursion relations associated with holomorphic all-line supershifts. This family of shifts similarly depends on
 reference parameters $|X]$ and $\eta_{\Y a}$, which then results in the dependence of the super MHV vertex expansion on these parameters.
 The ordinary MHV vertex expansion corresponds to
 the special case $\eta_{\Y a}=0$, but certain non-trivial choices
 for $\eta_{\Y a}$ can significantly reduce the number of diagrams contributing to the expansion and thus simplify the task of computing amplitudes.

The efficient computation of on-shell tree amplitudes
of $\cn=4$ SYM theory
has various applications. For example, tree amplitudes are an important ingredient for the computation
of loop amplitudes using (generalized) unitarity cuts\cite{Bern:1994zx,Bern:1994cg, Bern:1996ja,Bern:1997nh,Bern:2004ky,Bern:2005iz,Bern:2007ct,Cachazo:2008dx,Forde:2007mi,Bern:2007hh}.
The scattering amplitudes of $\cn=4$ SYM theory
are of particular interest, as the AdS/CFT correspondence permits insights into their strong coupling behavior (see \eg
\cite{Alday:2008yw} and references therein).

Another application
is the computation of on-shell tree amplitudes of $\cn=8$ supergravity, which
can be expressed
in terms of $\cn=4$ SYM amplitudes
through the KLT relations \cite{Kawai:1985xq}.
They then also play an important role in the
study of $\cn=8$ supergravity at loop level \cite{Bern:2008pv}.
Loop amplitudes in $\cn=8$ supergravity exhibit surprising properties
\cite{Bern:2005bb,BjerrumBohr:2006yw,BjerrumBohr:2008ji,BjerrumBohr:2008vc,ArkaniHamed:2008gz},
and
their UV behavior has recently been under intense investigation due to the
possible
perturbative finiteness of the theory \cite{Green:2006gt,Green:2006yu,Bern:2006kd,Bern:2007hh,Bern:2007xj,Bern:2008pv,Naculich:2008ew,Brandhuber:2008tf,
Bern:2009kf
}.

KLT relations can be used to relate not only
the on-shell tree amplitudes of $\cn=8$ supergravity and $\cn=4$ SYM,
but their generating functions as well.
Very recently, this was carried out in~\cite{Drummond:2009ge} using the generating functions~\cite{Drummond:2008cr} based on dual superconformal
symmetry \cite{Drummond:2006rz,Alday:2007hr,Drummond:2007aua, Drummond:2007cf,Drummond:2007au,Drummond:2008vq,Alday:2007he,Brandhuber:2008pf,Ricci:2007eq,Beisert:2008iq,McGreevy:2008zy,Berkovits:2008ic,Drummond:2008bq, Drummond:2009fd},
and using the KLT relations in the form of ref.~\cite{Elvang:2007sg}.
The final expression for the resulting $\cn=8$ supergravity generating function contained far fewer terms than naively expected~\cite{Drummond:2009ge}. It would be interesting to see whether a similar simplification occurs when the  generating function of the $\cn=4$ (super) MHV vertex expansion is used to determine an $\cn=8$ supergravity generating function via KLT.

It would also be interesting to find generating functions for $\cn=8$ supergravity amplitudes directly from $\cn=8$ recursion relations. To derive recursion relations, one needs to determine shifts (or supershifts) under which an amplitude (or generating function) vanishes as the deformation parameter $z$ is taken to infinity.  $\cn=8$ supergravity amplitudes generically do not vanish under holomorphic shifts when the number $n$ of external lines becomes large. In fact, under holomorphic shifts, amplitudes go as $z^{n-\ell}$ for some integer $\ell$.
For example, pure-graviton NMHV amplitudes go as $z^{n-12}$ under a holomorphic shift of the three negative helicity graviton
lines~\cite{Bianchi:2008pu,Benincasa:2007qj}.
The MHV vertex expansion for gravity~\cite{BjerrumBohr:2005jr}
is then not valid for graviton NMHV amplitudes with $n\geq 12$ external lines.
The falloff becomes even worse for more general external states,
and the MHV vertex expansion in $\cn=8$ supergravity
has not even been established for general $5$-point NMHV amplitudes,
and has been shown to
fail for certain scalar amplitudes at the $6$-point level.

In this paper, we found that holomorphic all-line supershifts with
suitably chosen shift parameters yield $1/z^{k+4}$ suppression
for anti-MHV amplitudes in $\cn=4$ SYM theory.
This immediately implies, via the KLT relations,
that $\cn=8$ supergravity anti-MHV generating functions
with $n>4$ external legs go at least as $z^{n-11-2k}=z^{-n-3}$
under a suitable  holomorphic all-line supershift.
A valid recursion relation, namely the super MHV vertex expansion
for $\cn=8$ supergravity, can thus be derived at the anti-MHV level.\footnote{
To see this,
observe
that all-line supershift recursion relations
express an $n$-point anti-MHV generating function purely
in terms of lower-point anti-MHV generating functions and
$3$- or $4$-point MHV generating functions.
A super MHV vertex expansion for arbitrary
anti-MHV generating functions in $\cn=8$
supergravity can thus be established inductively in $n$. }
It would be interesting to determine the precise form of this
expansion, and to study the sum rules implied
by $\eta_{\Y a}$
independence.\footnote{Other interesting sum rules
for $\cn=8$ supergravity were recently studied in~\cite{Spradlin:2008bu}.}
It would
be particularly
interesting to see whether the improved falloff of $1/z^{k+4}$
in $\cn=4$ SYM theory, and thus the improved falloff of
at least $z^{n-11-2k}$ in $\cn=8$ supergravity, can be generalized
beyond the anti-MHV level.
If so, the validity of a super MHV vertex expansion for $\cn=8$ supergravity
could also be extended beyond the anti-MHV level.

In this paper, we presented the super MHV vertex expansion as an on-shell recursion relation associated with a complex shift.
Various off-shell approaches, however, have also proved useful to gain
insights into the ordinary MHV vertex
expansion \cite{Gorsky:2005sf,Mansfield:2005yd,Vaman:2005dt,Feng:2006yy,Brandhuber:2006bf,Boels:2006ir,Boels:2007qn,Brandhuber:2007vm,Ananth:2007zy,Mason:2008jy}.
It would be interesting to see whether the super MHV vertex expansion has a natural interpretation in an off-shell framework.

\section*{Acknowledgments}

We are grateful to Henriette Elvang and Dan Freedman
for their involvement and insights
in the early stages of this project,
and for valuable comments on a draft version of this paper.
We also thank Marcus Spradlin for prompting us to derive
an analytic expression for the entries of table \ref{tabeliminate}.
SGN thanks the organizers of the ``Strings and Gauge Theories Workshop''
at the Michigan Center for Theoretical Physics
in September 2008
for providing the relaxed and stimulating environment
in which this work was initiated.
MK is supported by the US  Department of Energy through
cooperative research agreement DE-FG0205ER41360.
SGN's research is supported in part by the
NSF under grant PHY-0756518.

\bigskip

\appendix
\section{Large-$z$ falloff under holomorphic all-line supershifts}
\subsection{$\cf^\text{N$^k$MHV}_n\sim 1/z^k$  using the super BCFW recursion relations}\label{appBCFW}
In this appendix we outline the derivation of the $1/z^k$ falloff of N$^k$MHV generating functions under holomorphic all-line supershifts using the
supersymmetric generalization of the BCFW recursion relation
\cite{NimaParis,Brandhuber:2008pf,ArkaniHamed:2008gz,Drummond:2008cr}.
This generalizes the derivation of ref.~\cite{Kiermaier:2008vz}
from ordinary all-line shifts to all-line supershifts.

The $1/z^k$ falloff
of the N$^k$MHV generating function $\FNkMHV$ under an ordinary all-line shift
was derived
in ref.~\cite{Kiermaier:2008vz}
 by recursively studying the
 behavior of the
BCFW representation of N$^k$MHV amplitudes.
The inductive argument presented there relied on three facts:
\begin{enumerate}
  \item Each $\cn=4$ SYM amplitude admits at least one valid BCFW recursion relation\cite{ArkaniHamed:2008yf,Cheung:2008dn,Elvang:2008na}\,.
  \item An all-line shift on an amplitude acts, to leading order in $z$, as an all-line shift on the subamplitudes of each diagram in its BCFW representation.
  \item The falloff of $1/z^k$ is valid for all MHV and anti-MHV amplitudes.
\end{enumerate}
As a generalization of this argument, we
now use the super BCFW recursion relations
to study the behavior of generating functions under all-line supershifts. It suffices to repeat steps (1)--(3) for this case, and we will now briefly outline how this is done.

As shown in ref.~\cite{ArkaniHamed:2008gz},
all generating functions vanish at large $z$ under a super BCFW
shift (\ref{pqsuper}) for
any choice of two lines $p$ and $q$, and thus admit a valid
super BCFW recursion relation. This establishes the analog of (1).

For (2), we need to show that an all-line supershift
on the entire generating function acts,
to leading order in $z$, as an all-line supershift on the subamplitudes
of each diagram in the super BCFW representation.
Notice that the kinematics
are unaltered compared to the ordinary all-line shift, and the analysis
of ref.~\cite{Kiermaier:2008vz} thus establishes that the angle and square
brackets of the super BCFW subamplitudes are subject to
a holomorphic all-line supershift.
In particular, it follows from ref.~\cite{Kiermaier:2008vz}
that the square spinor associated with
the internal line of momentum $P_\a$
in a $[p,q\>$ super BCFW recursion relation shifts as
\be
|P_\a]  ~~\to~~ z c_{P_\a} |X] + \co(1),
 \qquad\text{with}\qquad
c_{P_\a} = \sum_{i\in \a} c_i\,
 \frac{ \<p i\> }{\< p q \> }
\label{internalshift}
\ee
under the all-line supershift.
It remains to analyze the dependence of the super BCFW subamplitudes
on the Grassmann variables.
We recall from \eqn{invariantdelta}
that all-line supershifts are designed to leave the argument of the overall
$\delta^{(8)}$ in the generating function
invariant.
For any super BCFW diagram characterized
by an internal line of momentum $P_\a\,$,
it is easy to show that
the change in the argument of the $\delta^{(8)}$ of either of
its subamplitudes is $O(1)$ under the all-line supershift
provided that the Grassmann variable associated with the internal
line undergoes the shift
\begin{equation}\label{cov}
\eta_{P_\a a}~~\to~~
\eta_{P_\a a}+z\,c_{P_\a}\eta_{\Y  a}\,,
\end{equation}
where $c_{P_\a}$ is defined in \eqn{internalshift}.
The Grassmann variables $\eta_{P_\a a}$ are integrated over to carry out the intermediate state sum in the super BCFW diagram. Therefore, the shift~(\ref{cov}) can be implemented as a change of variables in the Grassmann integral over $\eta_{P_\a a}$.
After this change of variables,
an all-line supershift on the whole amplitude
manifestly acts,
to leading order, as an all-line supershift on the super
BCFW subamplitudes. This establishes the analog of (2).

The analog of (3), \ie the $1/z^k$ behavior of MHV and anti-MHV generating functions under all-line supershifts, was established in
 section~\ref{secantimhv}.
The inductive argument of ref.~\cite{Kiermaier:2008vz} for ordinary
all-line shifts thus carries over to the case of supershifts.

\subsection{$\cf^\text{N$^k$MHV}_n \sim 1/z^k$  using the super MHV vertex expansion}\label{appconsistency}

In this paper, we have
seen the super MHV vertex expansion emerge from two different
approaches:
first, in section~\ref{secsuperMHV}, as a
supersymmetry transformation of the ordinary MHV vertex expansion,
and
second, in
 section~\ref{secsuperfromall},
as the recursion relation implied by all-line supershifts.
In this appendix we
close the loop and show that the super MHV vertex expansion immediately implies the $1/z^k$ falloff of N$^k$MHV generating functions
under all-line supershifts.
This serves as another consistency check on our result.

Consider the action of an all-line
supershift~(\ref{holosuper})
on the $n$-point N$^k$MHV generating function.
We choose to represent this generating function as the super MHV vertex
expansion  (\ref{complete}) with the reference parameters $|X]$ and $\eta_{\Y a}$ chosen to coincide with
those of the supershift.
Inspecting \eqn{complete}, we find that all $k$ propagators shift as
$1/\hat P^2_{\alpha_A}\sim 1/z$,
giving $1/z^k$ suppression.
The cyclic factors in the denominator are invariant because the CSW spinors $|P_{\alpha_A}\>$ are invariant by \eqn{unaffected}.
The spin factors
$\sum_{i\in\alpha_A}\<i\,P_{\alpha_A}\>\eta_{ia}\!+\!P_{\alpha_A}^2\eta_{\Y a}$
in the numerator are also invariant, by the identity~(\ref{crucial}). We thus find $1/z^k$ suppression diagram-by-diagram in the super MHV vertex expansion, and conclude that N$^k$MHV generating functions fall off at least as  $1/z^k$ under all-line supershifts.
As we saw in section~\ref{secantimhv},
the falloff can be even stronger
for certain amplitudes and shifts, but unfortunately the super MHV vertex expansion is not sensitive to
this stronger falloff, which arises through cancellations between diagrams.

This argument was, not surprisingly, simpler than the derivations
of the $1/z^k$ falloff from the ordinary MHV vertex expansion
and the super BCFW recursion relation
that we presented in section~\ref{seclargez}
and appendix~\ref{appBCFW}, respectively.
Just as the $1/z^k$ falloff under ordinary all-line shifts was naturally shown from the ordinary MHV vertex expansion with coinciding reference spinor in
ref.~\cite{Kiermaier:2008vz},
we have now done the same for all-line supershifts using the super MHV vertex expansion
with coinciding reference spinor and coinciding reference Grassmann parameters.

\providecommand{\href}[2]{#2}\begingroup\raggedright\endgroup

\end{document}